\pdfoutput=1

\documentclass[twocolumn,final]{svjour3}

\smartqed

\usepackage{graphicx,amssymb}
\usepackage{epsf,psfig}
\usepackage{txfonts}
\usepackage{natbib}

\journalname{Nonlinear Dynamics}

\begin{document}

\title{Determining the nature of orbits in disk galaxies with non spherical nuclei}

\author{Euaggelos E. Zotos \and Nicolaos D. Caranicolas}

\institute{Department of Physics, \\
Section of Astrophysics, Astronomy \& Mechanics, \\
Aristotle University of Thessaloniki, \\
GR-541 24, Thessaloniki, Greece \\
Corresponding author's email: {evzotos@physics.auth.gr}
}

\date{Received: 4 September 2013 / Accepted: 19 October 2013 / Published online: 14 November 2013}

\titlerunning{Determining the nature of orbits in disk galaxies with non spherical nuclei}

\authorrunning{E. E. Zotos \& N. D. Caranicolas}

\maketitle

\begin{abstract}

We investigate the regular or chaotic nature of orbits of stars moving in the meridional plane $(R,z)$ of an axially symmetric galactic model with a flat disk and a central, non spherical and massive nucleus. In particular, we study the influence of the flattening parameter of the central nucleus on the nature of orbits, by computing in each case the percentage of chaotic orbits, as well as the percentages of orbits of the main regular families. In an attempt to maximize the accuracy of our results upon distinguishing between regular and chaotic motion, we use both the Fast Lyapunov Indicator (FLI) and the Smaller ALingment Index (SALI) methods to extensive samples of orbits obtained by integrating numerically the equations of motion as well as the variational equations. Moreover, a technique which is based mainly on the field of spectral dynamics that utilizes the Fourier transform of the time series of each coordinate is used for identifying the various families of regular orbits and also to recognize the secondary resonances that bifurcate from them. Varying the value of the flattening parameter, we study three different cases: (i) the case where we have a prolate nucleus (ii) the case where the central nucleus is spherical and (iii) the case where an oblate massive nucleus is present. Furthermore, we present some additional findings regarding the reliability of short time (fast) chaos indicators, as well as a new method to define the threshold between chaos and regularity for both FLI and SALI, by using them simultaneously. Comparison with early related work is also made.


\end{abstract}

\section{Introduction}
\label{intro}

Knowing the dynamical properties and the overall orbital structure of galaxies is an issue of paramount importance. Therefore, scientists in an attempt to explore and interpret their structure, they usually build and apply galactic dynamical models, which in most cases are mathematical expressions giving either the potential or the mass density of the galaxy, as a function of the radius $R$ from the center. The reader can find a variety of interesting models, describing motion in galaxies in [\citealp{BT08}]. Moreover, potential density pairs for galaxies were also presented by [\citealp{VL05}]. Over the last years, several types of galactic models have been proposed in an attempt to model the orbital properties in axially symmetric systems. A simple yet realistic axisymmetric logarithmic potential was introduced in [\citealp{Bin81}] for the description of galactic haloes at which the mass density drops like $R^{-2}$ [\citealp{E93}]. However, the most well-known model for cold dark matter (CDM) haloes is the flattened cuspy NFW model [\citealp{NFW96},\citealp{NFW97}], where the density at large radii falls like $R^{-1}$ . This model being self-consistent has a major advantage and that's why it is mainly used for conducting $N$-body simulations.

In order to obtain the mass profiles of galaxies, we have to use dynamical models describing the main properties of the galaxies. These models can be generated by deploying two main techniques: (i) using superposition of libraries of orbits (e.g., [\citealp{G03},\citealp{TSB04},\citealp{C06}]) or (ii) using distributions functions (e.g., [\citealp{DBVZ96},\citealp{GJSB98},\citealp{KSGB00}]). In the literature there are also other more specialized dynamical models combining kinematic and photometric data. For instance, axially symmetric Schwarzschild models were used by [\citealp{B06}], while [\citealp{H08}] used Jeans models in order to fit observational data in the X-ray potential introduced by [\citealp{HBG06}]. Furthermore, axisymmetric Schwarzschild models were also used by [\citealp{SG10}] to fit data derived from the Hubble Space Telescope (HST).

Over the last years, Schwarzschild's superposition method [\citealp{S79}] has been heavily utilized by several researchers (e.g., [\citealp{RZCMC97},\citealp{G03},\citealp{TSB04},\citealp{VME04},\citealp{KCEMd05},\citealp{TSB05}]) in order to model dark matter distributions in elliptical galaxies therefore, we deem it is necessary to recall and describe briefly in a few words the basic points of this interesting method. Initially, $N$ closed cells define the configuration space, while $K$ orbits extracted from a given mass distribution construct the phase space. Then, integrating numerically the equations of motion, we calculate the amount of time spent by each particular orbit in every cell. Thus, the mass of each cell is directly proportional to the total sum of the stay times of orbits in every cell. Using this technique, we manage to compute the unknown weights of the orbits, assuming they are not negative.

Spherical analytical models describing the motion of stars in galaxies were studied by (e.g., [\citealp{D12},\citealp{RDZ05},\citealp{Z96}]). Moreover, interesting axially symmetric galactic models were presented by [\citealp{CZMR99}]. Recently, [\citealp{Z11}] used data derived from rotation curves of real galaxies, in order to construct a new axially symmetric model describing star motion in both elliptical and disk galaxy systems.

Of particular interest, are the so-called composite galactic dynamical models. In those models the potential has several terms each one describing a distinct component of the stellar system. Such a dynamical model with four components, that is a disk, a nucleus, a bulge and a dark halo was studied by [\citealp{C97}]. A new composite mass model describing motion in axially symmetric galaxies with dark matter was recently presented and studied by [\citealp{C12}]. Composite axially symmetric galaxy models describing the orbital motion in our galaxy were also studied by [\citealp{Bin12}]. In these models, the gravitational potential is generated by three superposed disks: one representing the gas layer, one the thin disk and one representing the thick disk.

In a previous work [\citealp{ZCar13a}], we introduced a new dynamical model describing three-dimensional motion in non axially symmetric galaxies. This model covers a wide range of galaxies from a disk system to an elliptical galaxy by suitably choosing the dynamical parameters. It was found, that the parameter which describes and controls the deviation from axially symmetry is indeed very influential both in the disk and the elliptical galaxy models. In the same vein, we proposed in [\citealp{CZ13},\citealp{ZCar13b}] two analytical models describing the main body of a disk or an elliptical galaxy which contains dark matter. In the first model [\citealp{CZ13}] for the main body of the galaxy we used a mass potential, while in the second model [\citealp{ZCar13b}] it was described by a logarithmic potential. The fractional portion of the dark matter in the main body of the galaxy is regulated by a parameter and our numerical calculations suggested that in both models this parameter plays a key role to the overall orbital structure of the system.

Therefore, taking into account all the above there is no doubt, that modelling galaxies is an issue of great importance. On this basis, it seems of particular interest to build an analytical axially symmetric dynamical model describing the motion of stars in disk galaxies with non spherical and massive nuclei and also explore, how the deviation from axially symmetry of the central nucleus, being prolate or oblate, affects the regular or chaotic character of orbits as well as the behavior of the different families of orbits. Similar research preformed in [\citealp{HN90},\citealp{HPN93},\citealp{Z12a}], where the effects of a central, spherical mass component in a galaxy were investigated.

The present article is organized as follows: In Section \ref{galmod} we present in detail the structure and the properties of our gravitational galactic model. In Section \ref{compmeth} we describe the computational methods we used in order to determine the character of orbits. In the following Section, we investigate how the flattening parameter of the central non spherical nucleus influences the character of the orbits. The next Section is devoted to some additional findings of our investigation regarding the reliability of chaos indicators. Our paper ends with Section \ref{disc}, where the discussion and the conclusions of this research are presented.

\section{Presentation and properties of the galactic model}
\label{galmod}

In this work, we use an axially symmetric disk galaxy model with an additional non spherical nucleus in order to investigate how the deviation from spherical symmetry of the central nucleus influences the regular or chaotic character of orbits in the meridional plane $(R,z)$. We shall use the usual cylindrical coordinates $(R,\phi,z)$, where $z$ is the axis of symmetry.

The total potential $V(R,z)$ in our model is the sum of a disk potential $V_{\rm d}$ and a central non spherical component $V_{\rm n}$. The first part is a generalization of the Miyamoto-Nagai potential [\citealp{MN75}] (see also [\citealp{CI87},\citealp{CI91}])
\begin{equation}
V_{\rm d}(R,z) = \frac{-G M_{\rm d}}{\sqrt{b^2 + R^2 + \left(a + \sqrt{h^2 + z^2}\right)^2}}.
\label{Vd}
\end{equation}
Here $G$ is the gravitational constant, $M_{\rm d}$ is the mass of the disk, $b$ represents the core radius of the disk-halo, while the parameters $\alpha$ and $h$ correspond to the horizontal and vertical scale length of the disk respectively.

For the description of the central non spherical nucleus we introduce the following potential
\begin{equation}
V_{\rm n}(R,z) = \frac{-G M_{\rm n}}{c + \sqrt{R^2 + \beta z^2}},
\label{Vn}
\end{equation}
where $M_{\rm n}$ is the mass of the nucleus, $c$ is the scale length of the nucleus (acting also as a softening parameter), while $\beta$ is the flattening parameter. This model is somehow a combination of two well known galaxy models: (a) the Plummer model [\citealp{P11}] and (b) the Hernquist's model [\citealp{H91}]. At this point, we must make clear that Eq. (\ref{Vn}) is not intended to represent the potential of a black hole nor that of any other compact object, but just the potential of a dense and massive nucleus thus, we do not include relativistic effects.

We use a system of galactic units, where the unit of length is 1 kpc, the unit of mass is $2.325 \times 10^7 {\rm M}_\odot$ and the unit of time is $0.9778 \times 10^8$ yr. The velocity unit is 10 km/s, the unit of angular momentum (per unit mass) is 10 km kpc$^{-1}$ s$^{-1}$, while $G$ is equal to unity. Finally, the energy unit (per unit mass) is 100 km$^2$s$^{-2}$. In these units, the values of the involved parameters are: $M_{\rm d} = 7500$ (corresponding to 1.74 $\times$ $10^{11}$ M$_{\odot}$), $b = 6$, $\alpha = 3$, $h = 0.15$, $M_n = 400$ (corresponding to 9.3 $\times$ $10^{9}$ M$_{\odot}$) and $c = 0.25$. The values of the disk and the nucleus were chosen having in mind a Milky Way-type galaxy (e.g., [\citealp{AS91}]). The flattening parameter $\beta$, on the other hand, is treated as a parameter and its value varies in the interval $0.1 \leq \beta \leq 1.9$. The flattening parameter controls the particular shape of the central non spherical nucleus. In particular, when $0.1 \leq \beta < 1$ the nucleus is prolate, when $\beta = 1$ is spherical, while when $1 < \beta \leq 1.9$ is oblate.

\begin{figure}
\includegraphics[width=\hsize]{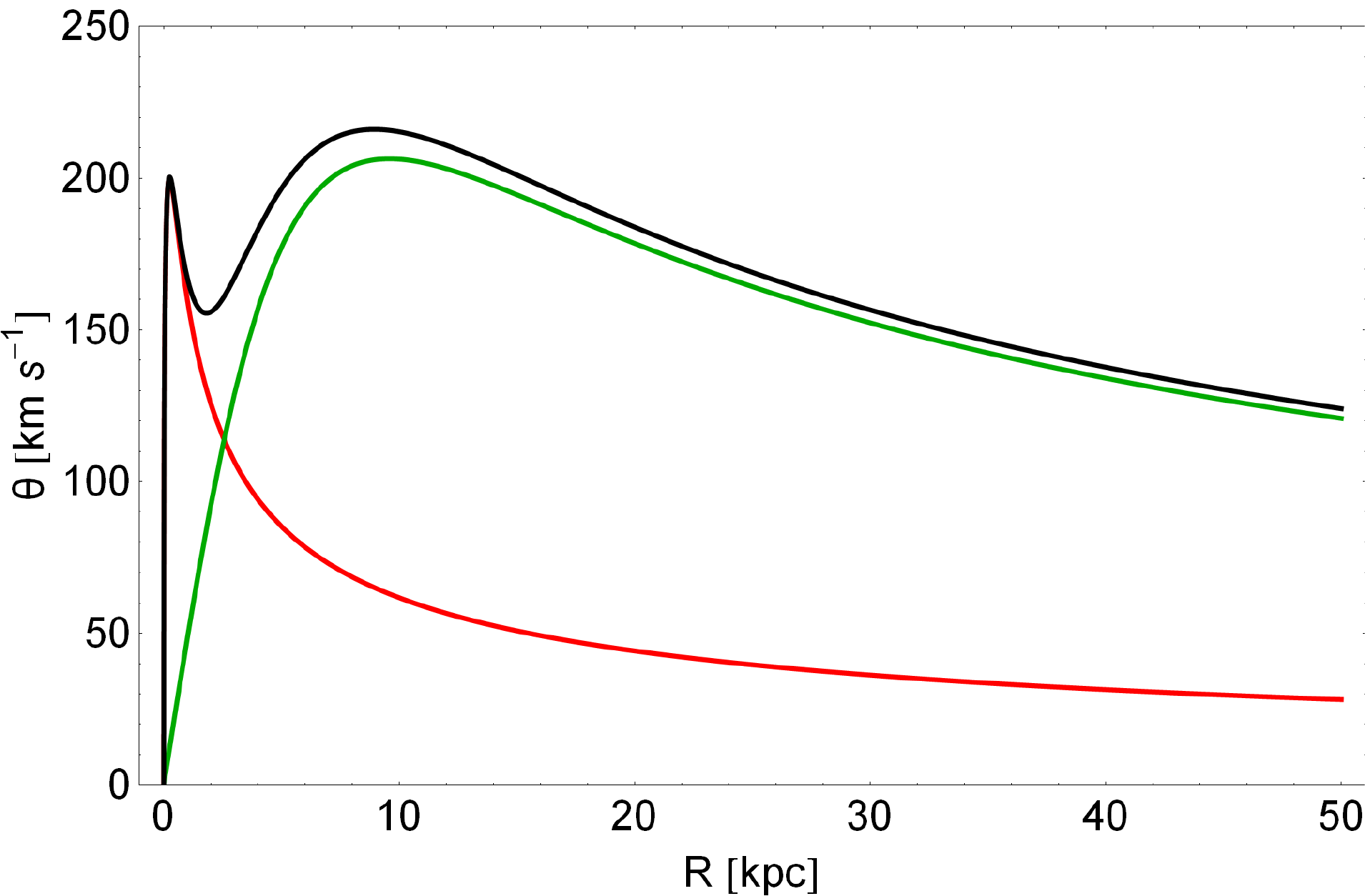}
\caption{A plot of total circular velocity $\theta(R)$ of the galactic model (black). We can also distinguish the contributions from the disk (green) and that of the central non spherical nucleus (red).}
\label{rotvel}
\end{figure}

\begin{figure*}
\centering
\resizebox{0.80\hsize}{!}{\includegraphics{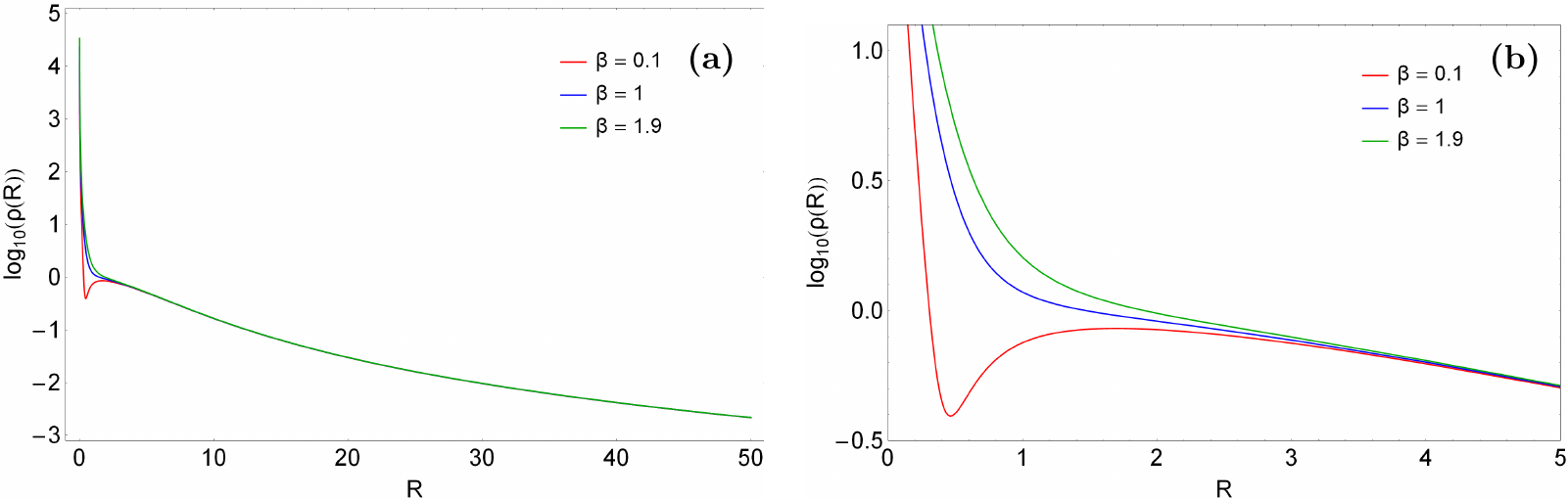}}
\caption{(a-left): Evolution of the mass density $\rho(R)$ in the galactic plane $(z = 0)$, as a function of the distance $R$ from the center for three different values of the flattening parameter $\beta$. (b-right): Details of Fig. \ref{denevol}a.}
\label{denevol}
\end{figure*}

\begin{figure}
\includegraphics[width=\hsize]{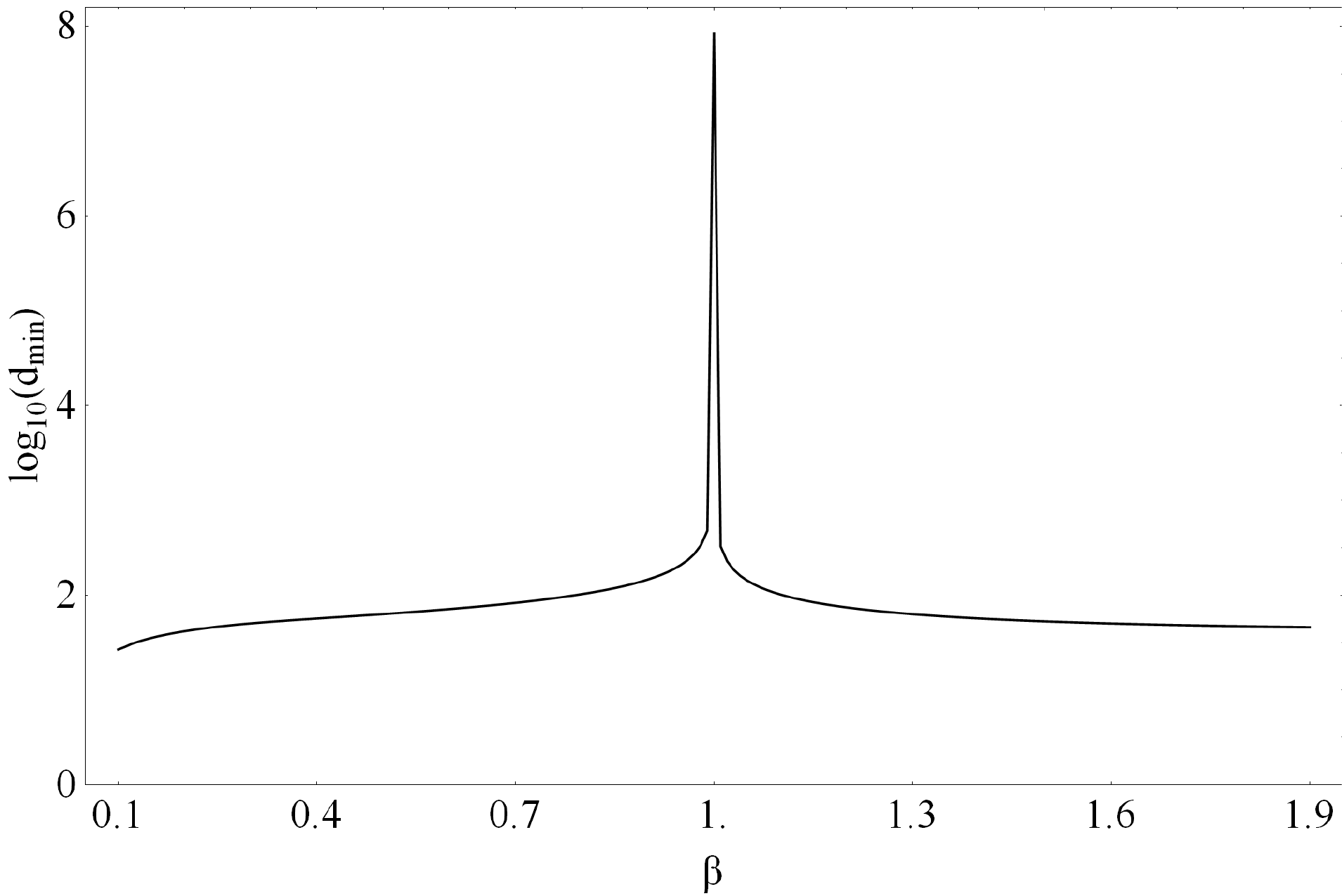}
\caption{The evolution of the minimum distance $d_{\rm min}$ where negative density appears for the first time, as a function of the flattening parameter $\beta$.}
\label{deneg}
\end{figure}

In disk galaxies the circular velocity in the galactic plane $z=0$,
\begin{equation}
\theta(R) = \sqrt{R\left|\frac{\partial V}{\partial R}\right|_{z=0}},
\label{cvel}
\end{equation}
is a physical quantity of great importance. Inserting the total potential $V(R,z)$ in Eq. (\ref{cvel}) we obtain
\begin{equation}
\theta(R) = R \sqrt{G\left(\frac{M_{\rm d}}{\left(b^2 + R^2 + (\alpha + h)^2\right)^{3/2}} +
\frac{M_{\rm n}}{R\left(R + c\right)^2}\right)}.
\label{cvelA}
\end{equation}
A plot of $\theta(R)$ as a function of the radius $R$ from the galactic center is presented in Fig. \ref{rotvel}, as a black curve. Furthermore, in the same plot the green curve is the contribution from the disk, while the red line corresponds to the contribution of the non spherical nucleus. It is seen, that at small radii from the galactic center, $R \leq 2.56$ kpc, the contribution from the central nucleus dominates, while at larger distances, $R > 2.56$ kpc, the disk contribution is the dominant factor. We also observe, the characteristic local minimum of the rotation curve due to the massive nucleus, which appears at small values of $R$, when fitting observed data to a galactic model [\citealp{GHBL10},\citealp{IWTS13}].

It is very useful to compute the mass density $\rho(R,z)$ derived from the total potential $V(R,z)$ using the Poisson's equation
\begin{eqnarray}
\rho(R,z) &=& \frac{1}{4 \pi G} \nabla^2 V(R,z) = \nonumber \\
&=& \frac{1}{4 \pi G} \left(\frac{\partial^2}{\partial R^2} + \frac{1}{R}\frac{\partial}{\partial R}
+ \frac{1}{R^2} \frac{\partial^2}{\partial \phi^2} + \frac{\partial^2}{\partial z^2}\right) V(R,z).
\label{dens}
\end{eqnarray}
Remember, that due to the axial symmetry the third term in Eq. (\ref{dens}) is zero. If we set $z = 0$ in Eq. (\ref{dens}) we obtain the mass density on the galactic plane which is
\begin{equation}
\rho(R) = \frac{1}{4\pi}\left(\rho_{\rm d} + \rho_{\rm n}\right),
\label{densR}
\end{equation}
where
\begin{eqnarray}
\rho_{\rm d} &=& \frac{M_{\rm d}\left(\alpha^3 + h\left(5\alpha^2 + 3b^2 + 3h^2\right) + \alpha \left(b^2 + R^2 + 7h^2\right)\right)}{h\left(b^2 + R^2 + \left(a + h\right)^2\right)^{5/2}}, \nonumber \\
\rho_{\rm n} &=& \frac{M_{\rm n}\left(R\left(\beta - 1\right) + c\left(\beta + 1\right)\right)}{R\left(R + c \right)^3}.
\label{densR2}
\end{eqnarray}
In the following Fig. \ref{denevol}(a-b) we present the evolution of $\rho(R)$ as a function of the of the radius $R$ from the galactic center, for three values of the flattening parameter $\beta$. In the same diagram, the red line correspond to $\beta = 0.1$, the blue curve to $\beta = 1$, while the green line corresponds to $\beta = 1.9$. It is evident, that the mass density decreases rapidly obtaining very low values with increasing radius $R$. We also observe, that when $R > 4$ kpc all three curves coincide, suggesting that the evolution of the mass density at large radii is the same regardless the particular value of the flattening parameter. According to our numerical calculations, at large galactocentric distances the non spherical nucleus is the dominant component therefore, the mass density should varies like $R^{-3}$.

Here we must clarify, that the mass density $\rho(R,z)$ in our galaxy model obtains negative values when the distance from the centre of the galaxy described by the model exceeds a minimum distance $d_{min}$, which strongly depends on the flattening parameter $\beta$. Fig. \ref{deneg} shows a plot of the evolution of $d_{\rm min}$ as a function of $\beta$. We see, that even at the extreme values of the flattening parameter, that is when $\beta = 0.1$ or $\beta = 1.9$, the first indication of negative density occurs only when $d_{\rm min} > 40$ kpc, that is almost at the theoretical boundaries of a real galaxy. On the other hand, when the central nucleus is spherical $\beta = 1$, the negative values of density tend to infinity, as expected. Moreover, we must point out, that our gravitational potential is truncated ar $R_{\rm max} = 30$ kpc for both reasons: (i) otherwise the total mass of the galaxy modeled by this potential would be infinite, which is obviously not physical and (ii) to avoid the existence of negative values of density.

Taking into account that the total potential $V(R,z)$ is axisymmetric, the $z$-component of the angular momentum $L_z$ is conserved. With this restriction, orbits can be described by means of the effective potential
\begin{equation}
V_{\rm eff}(R,z) = V(R,z) + \frac{L_z^2}{2R^2}.
\label{veff}
\end{equation}
The $L_z^2/(2R^2)$ term represents a centrifugal barrier; only orbits with sufficient small $L_z$ are allowed to pass near the axis of symmetry. The three-dimensional (3D) motion is thus effectively reduced to a two-dimensional (2D) motion in the meridional plane $(R,z)$, which rotates non-uniformly around the axis of symmetry according to
\begin{equation}
\dot{\phi} = \frac{L_z}{R^2},
\end{equation}
where of course the dot indicates derivative with respect to time.

The equations of motion on the meridional plane are
\begin{equation}
\ddot{R} = - \frac{\partial V_{\rm eff}}{\partial R}, \ \ \ \ddot{z} = - \frac{\partial V_{\rm eff}}{\partial z},
\label{eqmot}
\end{equation}
while the equations describing the evolution of a deviation vector $\delta {\bf{w}} = (\delta R, \delta z, \delta \dot{R}, \delta \dot{z})$ which joins the corresponding phase space points of two initially nearby orbits, needed for the calculation of the standard chaos indicators (the FLI and SALI in our case) are given by the variational equations
\begin{eqnarray}
\dot{(\delta R)} &=& \delta \dot{R}, \ \ \ \dot{(\delta z)} = \delta \dot{z}, \nonumber \\
(\dot{\delta \dot{R}}) &=&
- \frac{\partial^2 V_{\rm eff}}{\partial R^2} \delta R
- \frac{\partial^2 V_{\rm eff}}{\partial R \partial z}\delta z,
\nonumber \\
(\dot{\delta \dot{z}}) &=&
- \frac{\partial^2 V_{\rm eff}}{\partial z \partial R} \delta R
- \frac{\partial^2 V_{\rm eff}}{\partial z^2}\delta z.
\label{vareq}
\end{eqnarray}

Consequently, the corresponding Hamiltonian to the effective potential given in Eq. (\ref{veff}) can be written as
\begin{equation}
H = \frac{1}{2} \left(\dot{R}^2 + \dot{z}^2 \right) + V_{\rm eff}(R,z) = E,
\label{ham}
\end{equation}
where $\dot{R}$ and $\dot{z}$ are the momenta per unit mass, conjugate to $R$ and $z$ respectively, while $E$ is the numerical value of the Hamiltonian, which is conserved. Therefore, an orbit is restricted to the area in the meridional plane satisfying $E \geq V_{\rm eff}$.

\section{Computational methods}
\label{compmeth}

In our study, an issue of paramount importance is knowing whether an orbit is regular or chaotic. Several chaos indicators are available in the literature; we chose two of them, namely the Fast Lyapunov Indicator (FLI) and the Smaller ALignment index (SALI). To compute these indicators, we integrated, along with each orbit, its corresponding variational equations (\ref{vareq}) from unitary displacements in each of the Cartesian directions of the phase space $(R, z, \dot{R}, \dot{z})$ of the meridional plane.

\begin{figure*}
\centering
\resizebox{0.90\hsize}{!}{\includegraphics{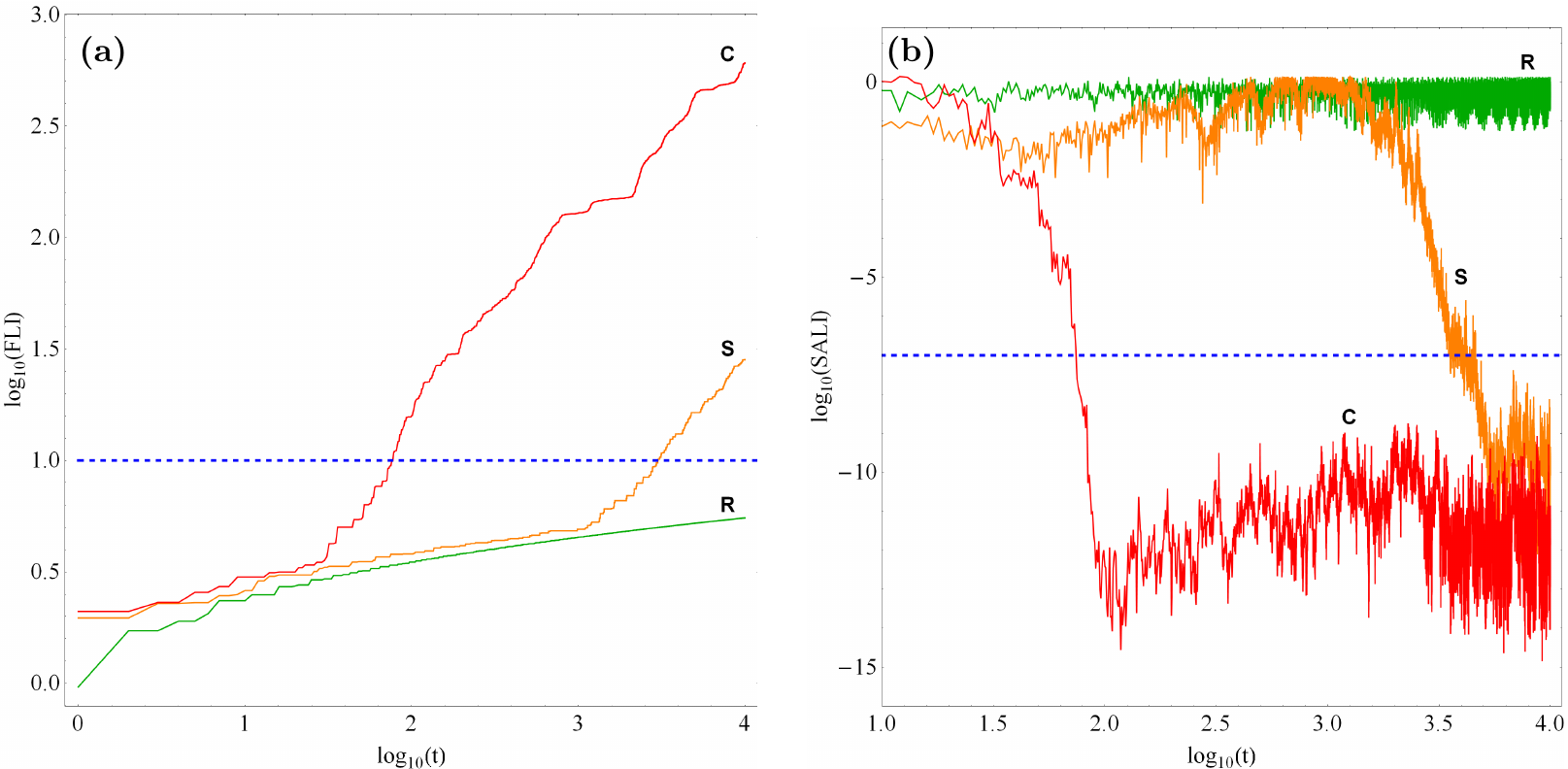}}
\caption{Time-evolution of (a-left): the FLI and (b-right): the SALI of a regular orbit (green color - R), a sticky orbit (orange color - S) and a chaotic orbit (red color - C) in our model for a time period of $10^4$ time units. The horizontal, blue, dashed lines corresponds to the threshold values of 10 (FLI) and $10^{-7}$ (SALI) which separate regular from chaotic motion. The FLI needs about 3000 time units of numerical integration in order to identify the true chaotic nature of the sticky orbit, while the SALI method requires about 3500 time units of numerical integration for the same purpose.}
\label{FSevol}
\end{figure*}

The FLI [\citealp{FGL97},\citealp{LF01}] is known as a very fast, reliable and effective tool, which is defined as
\begin{equation}
\rm FLI(t) = \log \| {\bf{w}(t)} \|, \ \ \ \ t \leq t_{max},
\end{equation}
where ${\bf{w}}(t)$ is a deviation vector. For distinguishing between regular and chaotic motion, we need to compute the FLI for a relatively short time interval of numerical integration $t_{max}$. In particular, we track simultaneously the time-evolution of the orbit itself as well as one deviation vector ${\bf{w}}(t)$ in order to compute the FLI. The variational equations (\ref{vareq}), as usual, are used for the evolution of the deviation vector. The particular time-evolution of the FLI allow us to distinguish between regular and chaotic motion as follows: when an orbit is regular the FLI exhibits a linear increase, while on the other hand, in the case of chaotic orbits the FLI increases super-exponentially. Unfortunately, this qualitative criterion is applicable only when someone wants to check the character of individual orbits by plotting and then inspecting by eye the evolution of FLI. Nevertheless, we can easily overtake this drawback by establishing a numerical threshold value, in order to quantify the results of FLI. A generally accepted threshold is the value 10. Hence, in order to decide whether an orbit is regular or chaotic, one may apply the usual method according to which we check after a certain and predefined time interval of numerical integration, if the value of FLI has become greater than the established threshold value. Therefore, if FLI $\geqslant 10$ the orbit is chaotic, while if FLI $ < 10$ the orbit is regular. The time-evolution of a regular (R) and a chaotic (C) orbit for a time period of $10^4$ time units is presented in Fig. \ref{FSevol}a.

For the computation of the SALI [\citealp{S01}], one has to follow simultaneously the time-evolution of the orbit itself as well as two deviation vectors ${\bf{w_1}}(t)$ and ${\bf{w_2}}(t)$, which initially point in two arbitrary directions. As usual, the evolution of the deviation vectors is given by the variational equations (\ref{vareq}). Moreover, taking into account that in this case we are only interested in the directions of these two deviation vectors, we normalize them at every time step of the numerical integration, thus keeping always their norm equal to unity. By applying this technique, we manage to control the exponential growth of the norm of the deviation vectors and we also avoid overflow malfunctions. Especially, when we deal with chaotic orbits the normalized deviation vectors point always to the very same direction and therefore, become equal or opposite in sign. Thus, the SALI can be obtained as
\begin{equation}
\rm SALI(t) = min(\| {\bf{w_1}}(t) + {\bf{w_2}}(t) \|, \| {\bf{w_1}}(t) - {\bf{w_2}}(t) \|),
\label{sali}
\end{equation}
where of course $t$ is the time, while $\| . \|$ denotes the Euclidean norm. By definition applies that $\rm SALI(t) \in [0, \sqrt{2}]$ and when SALI = 0 the two normalized deviation vectors are equal or opposite, nevertheless pointing to the same direction. It is the unique properties of the time-evolution of the SALI that allow us to distinguish between regular and chaotic motion as follows: in the case of regular orbits the SALI exhibits small fluctuations around a non-zero value, while on the other hand, in the case of chaotic orbits after a small transient period it tends exponentially to zero approaching the limit of the accuracy of the computer $(10^{-16})$. In Fig. \ref{FSevol}b we see the time-evolution of a regular (R) and a chaotic (C) orbit for a time period of $10^4$ time units. Generally speaking, two different initial deviation vectors become tangent to different directions on the torus, thus producing different sequences of vectors leading the SALI always to fluctuates around positive values. On the contrary, for chaotic orbits, any two initially different deviation vectors in time always tend to align in the direction defined by the maximal Lyapunov Characteristic Number (mLCN) (e.g., [\citealp{J91}]). Therefore, they either coincide with each other or become opposite, which forces the SALI to decrease rapidly to zero. Thus, we can exploit this completely different behaviour of the SALI in order to discriminate between regular and chaotic motion in Hamiltonian systems of any dimensionality.

In order to investigate the orbital properties (chaoticity or regularity) of the dynamical system, we need to establish some sample of initial conditions of orbits. The best approach, undoubtedly, would have been, if we could extract these sample of orbits from the distribution function of the model. Unfortunately, this is not available so, we followed another course of action. For determining the chaoticity of our models, we chose, for each set of values of the flattening parameter, a dense grid of initial conditions in the $(R,\dot{R})$ phase plane, regularly distributed in the area allowed by the value of the energy $E$. The points of the grid were separated 0.1 units in $R$ and 0.5 units in $\dot{R}$ direction. For each initial condition, we integrated the equations of motion (\ref{eqmot}) as well as the variational equations (\ref{vareq}) with a double precision Bulirsch-Stoer algorithm (e.g., [\citealp{PTVF92}]) with a small time step of order of $10^{-2}$, which is sufficient enough for the desired accuracy of our computations (i.e. our results practically do not change by halving the time step). In all cases, the energy integral (Eq. (\ref{ham})) was conserved better than one part in $10^{-10}$, although for most orbits it was better than one part in $10^{-11}$.

Each orbit was integrated numerically for a time interval of $10^4$ time units (10 billion yr), which corresponds to a time span of the order of hundreds of orbital periods but of the order of one Hubble time. The particular choice of the total integration time is an element of great importance, especially in the case of the so called ``sticky orbits" (i.e., chaotic orbits that behave as regular ones during long periods of time). A sticky orbit could be easily misclassified as regular by any chaos indicator\footnote{Generally, dynamical methods are broadly split into two types: (i) those based on the evolution of sets of deviation vectors in order to characterize an orbit and (ii) those based on the frequencies of the orbits which extract information about the nature of motion only through the basic orbital elements without the use of deviation vectors.}, if the total integration interval is too small, so that the orbit does not have enough time in order to reveal its true chaotic character. Thus, all the sets of the grids of the initial conditions of orbits were integrated, as we already said, for $10^4$ time units, thus avoiding sticky orbits with a stickiness at least of the order of a Hubble time. All the sticky orbits which do not show any signs of chaoticity for $10^4$ time units are counted as regular ones, since that vast sticky periods are completely out of scope of our research. A characteristic example of a sticky orbit (S) in our galactic system can be seen in Fig. \ref{FSevol}(a-b), where the horizontal, blue, dashed lines corresponds to the threshold values which separate regular from chaotic motion. We observe, that according to the usual threshold values the chaotic character of this particular sticky orbit is revealed after about 3000 time units using the FLI method and only after about 3500 time units when the SALI method is used.

To classify an orbit as regular or chaotic by using either the FLI or the SALI, a threshold value should be established separating both types of orbits. However, this is a delicate issue, as these thresholds are generally obtained by some statistical procedure, or even by eye inspection of plots of the time-evolution of indicators versus time. Besides, whereas the results of different chaos indicators agree in general, there are also the sticky orbits, that may be misclassified by one or another method depending of the threshold value used. We established the thresholds by taking advantage of our simultaneous computation of two chaos indicators: after the set of orbits of a grid has been integrated and the FLI and SALI were computed, we looked for those values of the thresholds that maximised the agreement in the classification of both methods. We found the threshold values which leave less than 1\% of orbits per grid differently classified by both methods. A more thorough description regarding the issue of the reliability of chaos indicators and the corresponding threshold values is given in Section \ref{fastIndi}. It is worth noticing, that in a previous work [\citealp{ZC13}] we used the exact same procedure dealing with mLCN and SALI, while on the other hand, [\citealp{KV05}] used a similar combination of mLCN and SALI, although they supplemented those indicators by adding the computation of the variation with time of the fundamental frequencies of the orbits, and they did not combine the indicators to establish their thresholds.

A first step towards the understanding of the overall behavior of our galactic system is knowing the regular or chaotic nature of orbits. Of particular interest, however, is also the distribution of regular orbits into different families. Therefore, once the orbits have been characterized as regular or chaotic, we then further classified the regular orbits into different families, by using the frequency analysis of [\citealp{CA98},\citealp{MCW05}]. Initially, [\citealp{BS82},\citealp{BS84}] proposed a technique, dubbed spectral dynamics, for this particular purpose. Later on, this method has been extended and improved by [\citealp{CA98}] and [\citealp{SN96}]. In a recent work, [\citealp{ZC13}] the algorithm was refined even further, so it can be used to classify orbits in the meridional plane. In general terms, this method calculates the Fourier transform of the coordinates of an orbit, identifies its peaks, extracts the corresponding frequencies and search for the fundamental frequencies and their possible resonances. Thus, we can easily identify the various families of regular orbits and also recognize the secondary resonances that bifurcate from them.

Before closing this Section, we would like to make a short note about the nomenclature of orbits. All the orbits of an axisymmetric potential are in fact three-dimensional (3D) loop orbits, i.e., orbits that rotate around the axis of symmetry always in the same direction. However, in dealing with the meridional plane the rotational motion is lost, so the path that the orbit follows onto this plane can take any shape, depending on the nature of the orbit. We will call an orbit according to its behaviour in the meridional plane. Thus, if for example an orbit is a rosette lying in the equatorial plane of the axisymmetric potential, it will be a linear orbit in the meridional plane, etc.

\begin{figure*}
\centering
\resizebox{0.9\hsize}{!}{\includegraphics{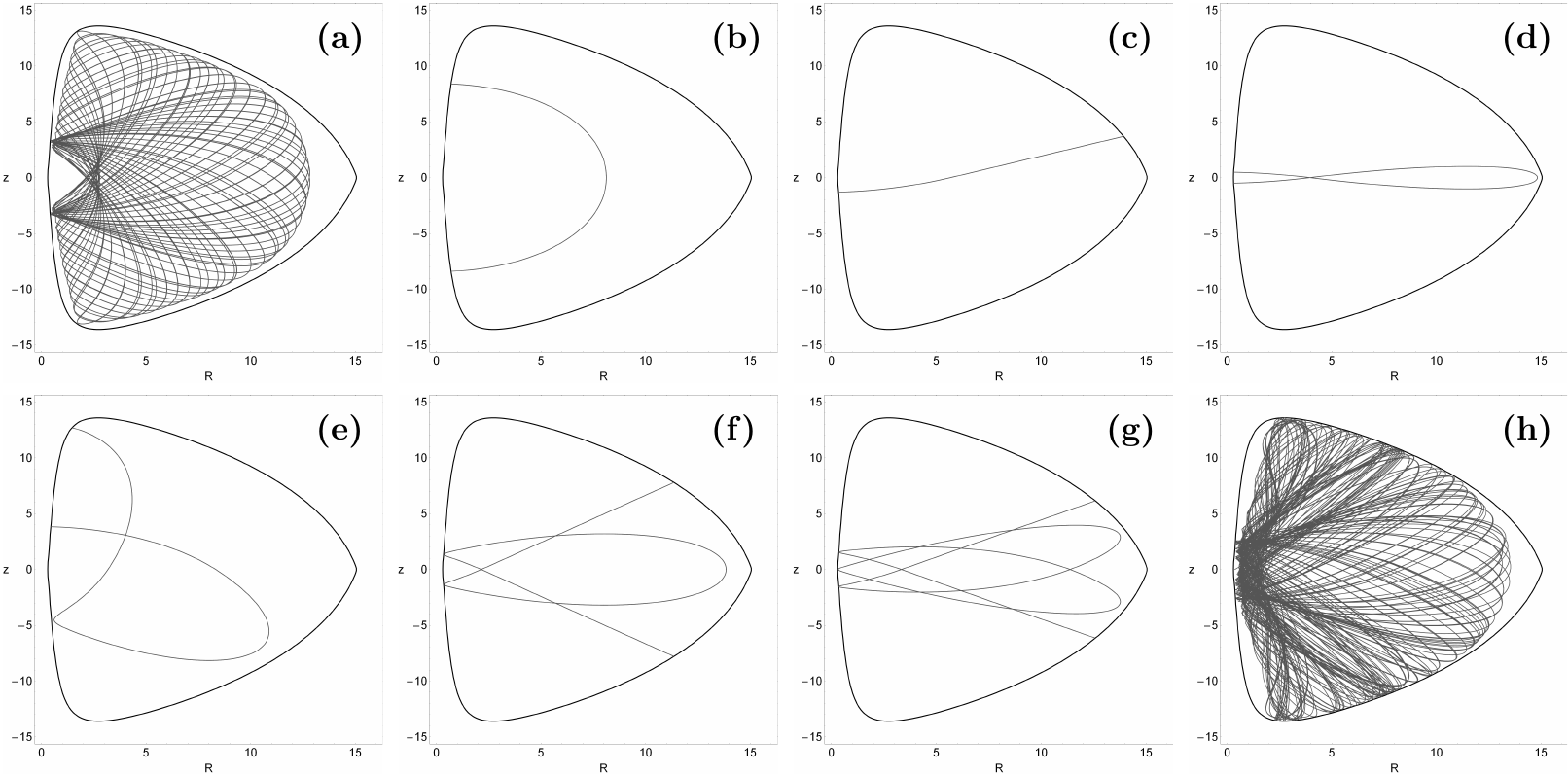}}
\caption{Orbit collection of the eight basic types of orbits in our galaxy model: (a) box orbit; (b) 2:1 banana-type orbit; (c) 1:1 linear orbit; (d) 2:3 boxlet orbit; (e) 3:2 boxlet orbit; (f) 4:3 boxlet orbit; (g) 6:5 boxlet orbit; (h) chaotic orbit.}
\label{orbs}
\end{figure*}

\begin{figure*}
\centering
\resizebox{0.9\hsize}{!}{\includegraphics{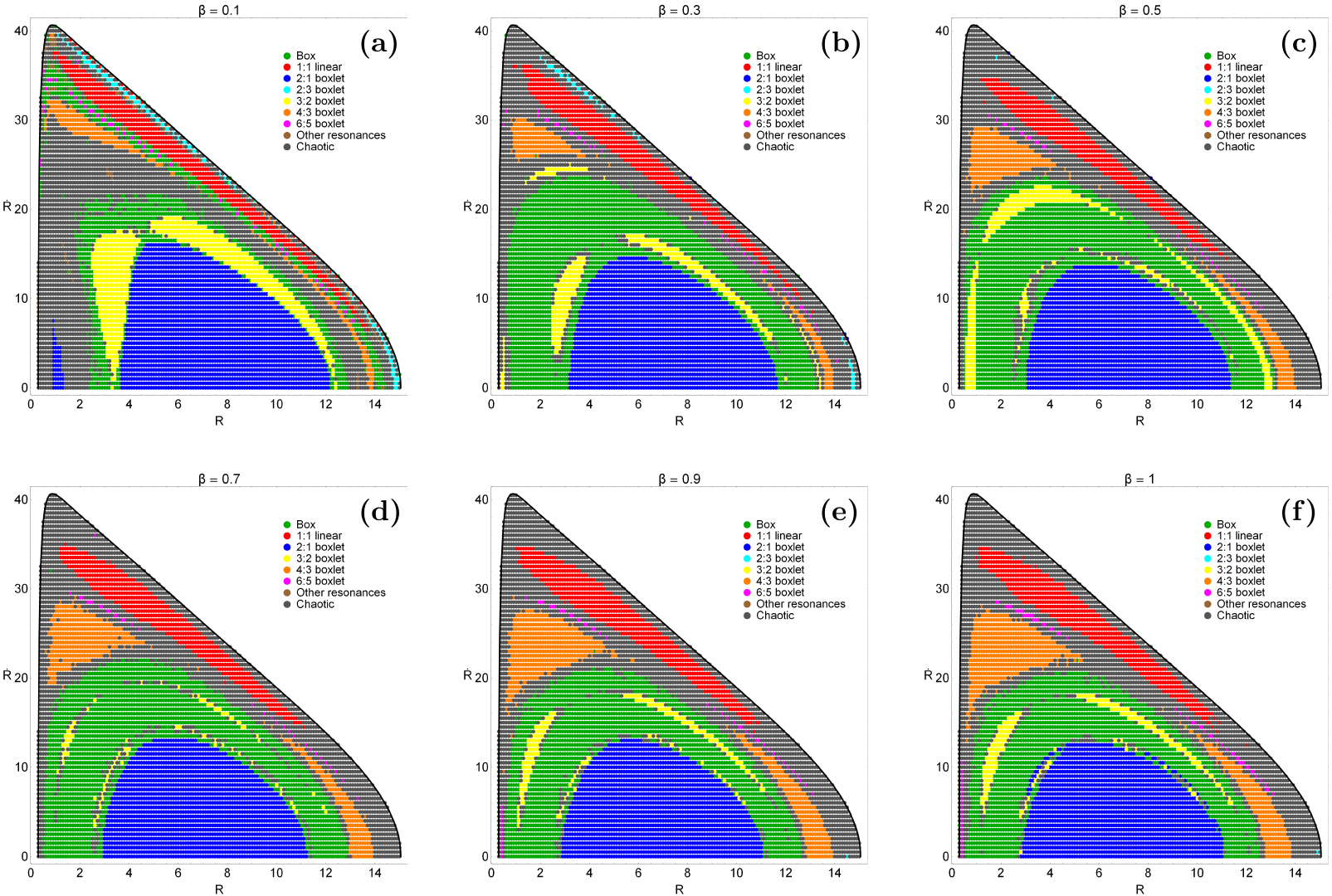}}
\caption{Orbital structure of the $(R,\dot{R})$ phase plane of the galaxy model for different values of the flattening parameter $\beta$, when $0.1 \leq \beta \leq 1$.}
\label{grdprl}
\end{figure*}

\begin{figure*}
\centering
\resizebox{0.9\hsize}{!}{\includegraphics{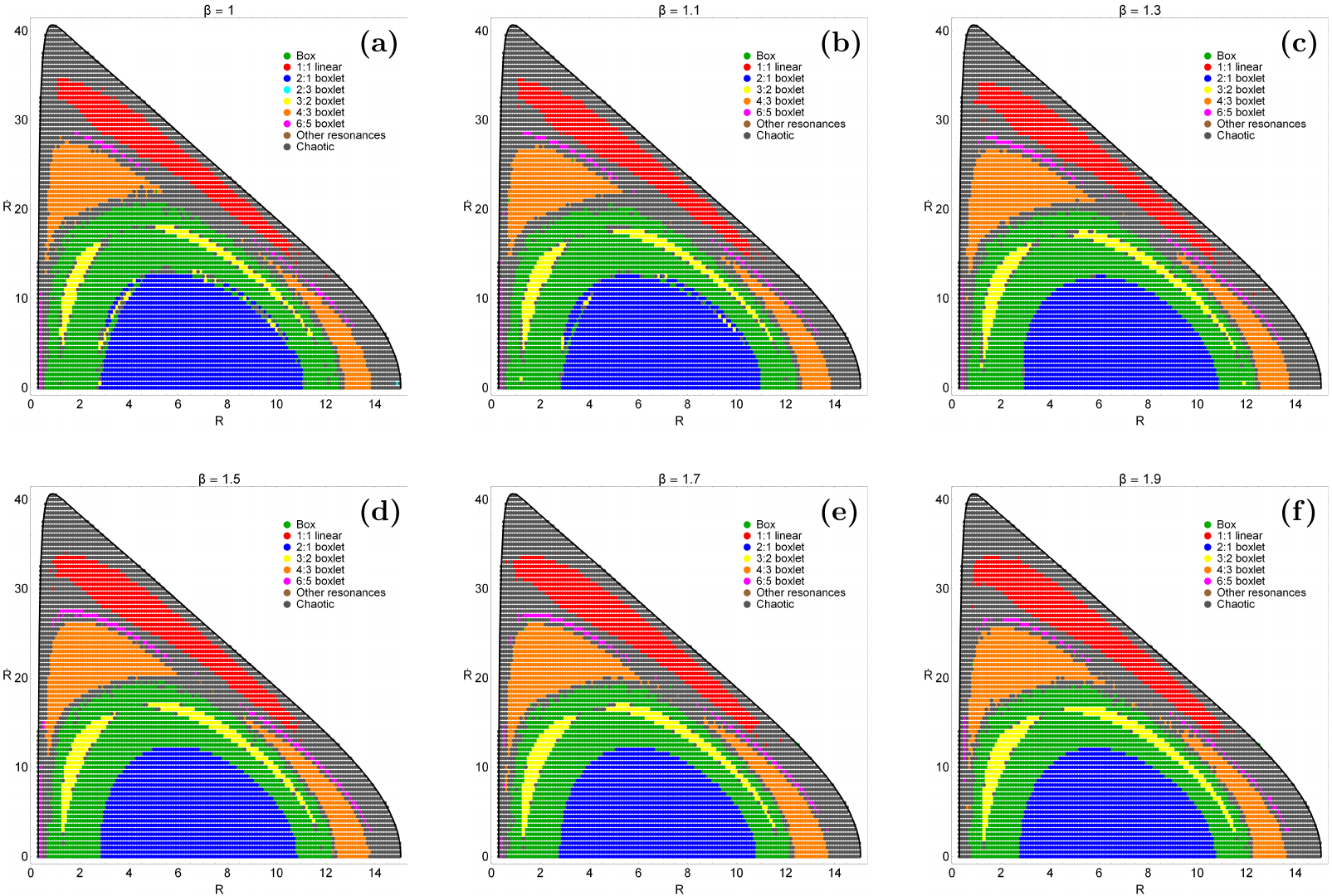}}
\caption{Orbital structure of the $(R,\dot{R})$ phase plane of the galaxy model for different values of the flattening parameter $\beta$, when $1 \leq \beta \leq 1.9$.}
\label{grdobl}
\end{figure*}

\section{Numerical results - Orbit classification}
\label{orbclas}

In this Section, we will integrate numerically several sets of orbits in an attempt to distinguish the regular or chaotic nature of motion. We use the initial conditions mentioned in Section \ref{compmeth} in order to construct the respective grids of initial conditions, taking always values inside the Zero Velocity Curve (ZVC) defined by
\begin{equation}
\frac{1}{2} \dot{R}^2 + V_{\rm eff}(R,0) = E.
\label{zvc}
\end{equation}
In all cases, the value of the angular momentum of the orbits is $L_z = 15$. We chose the energy level $E = -480$ which correspond to $R_{\rm max} \simeq 15$ kpc, where $R_{\rm max}$ is the maximum possible value of $R$ on the $(R,\dot{R})$ phase plane. Once the values of the parameters were chosen, we computed a set of initial conditions as described in Section \ref{compmeth} and integrated the corresponding orbits calculating the values of FLI and SALI indicators and then classifying the regular orbits into different families.

\begin{table}
\begin{center}
   \caption{Type and initial conditions of the galaxy model orbits shown in Figs. \ref{orbs}(a-h). In all cases, $z_0 = 0$ and $\dot{z_0}$ is found from the energy integral, Eq. (\ref{ham}), while $T_{\rm per}$ is the period of the resonant parent periodic orbits.}
   \label{table1}
   \setlength{\tabcolsep}{3.0pt}
   \begin{tabular}{@{}llccc}
      \hline
      Figure & Type & $R_0$ & $\dot{R_0}$ & $T_{\rm per}$  \\
      \hline
      \ref{orbs}a & box        &  2.75000000 &  0.00000000 &           - \\
      \ref{orbs}b & 2:1 banana &  8.12139235 &  0.00000000 &  2.52056954 \\
      \ref{orbs}c & 1:1 linear &  5.67183964 & 26.95576753 &  1.83037283 \\
      \ref{orbs}d & 2:3 boxlet & 14.82746221 &  0.00000000 &  3.68728638 \\
      \ref{orbs}e & 3:2 boxlet &  3.17090779 & 12.08482214 &  5.43688647 \\
      \ref{orbs}f & 4:3 boxlet & 13.84280496 &  0.00000000 &  7.21069423 \\
      \ref{orbs}g & 6:5 boxlet &  3.33014173 & 29.34720297 & 10.89547420 \\
      \ref{orbs}h & chaotic    &  0.36000000 &  0.00000000 &           - \\
      \hline
   \end{tabular}
\end{center}
\end{table}

Our investigation reveals, that in our axially symmetric galaxy model there are eight main types of orbits: (a) box orbits, (b) 1:1 linear orbits, (c) 2:1 banana-type orbits, (d) 2:3 resonant orbits, (e) 3:2 resonant orbits (f) 4:3 resonant orbits, (g) 6:5 resonant orbits and (h) chaotic orbits. Here we must emphasize, that every resonance $n:m$ is expressed in such a way that $m$ is equal to the total number of islands of invariant curves produced in the $(R,\dot{R})$ phase plane by the corresponding orbit. In Fig. \ref{orbs}(a-h) we present an example of each of the seven basic types of regular orbits, plus an example of a chaotic one. In all cases, we set $\beta = 0.1$. The orbits shown in Figs. \ref{orbs}a and \ref{orbs}h were computed until $t = 250$ time units, while all the parent periodic orbits were computed until one period has completed. The black thick curve circumscribing each orbit is the limiting curve in the meridional plane $(R,z)$ defined as $V_{\rm eff}(R,z) = E$. In Table \ref{table1} we give the type and the initial conditions for each of the depicted orbits, while for the resonant cases, the initial conditions and the period $T_{\rm per}$ correspond to the parent periodic orbits.

\begin{figure}
\includegraphics[width=\hsize]{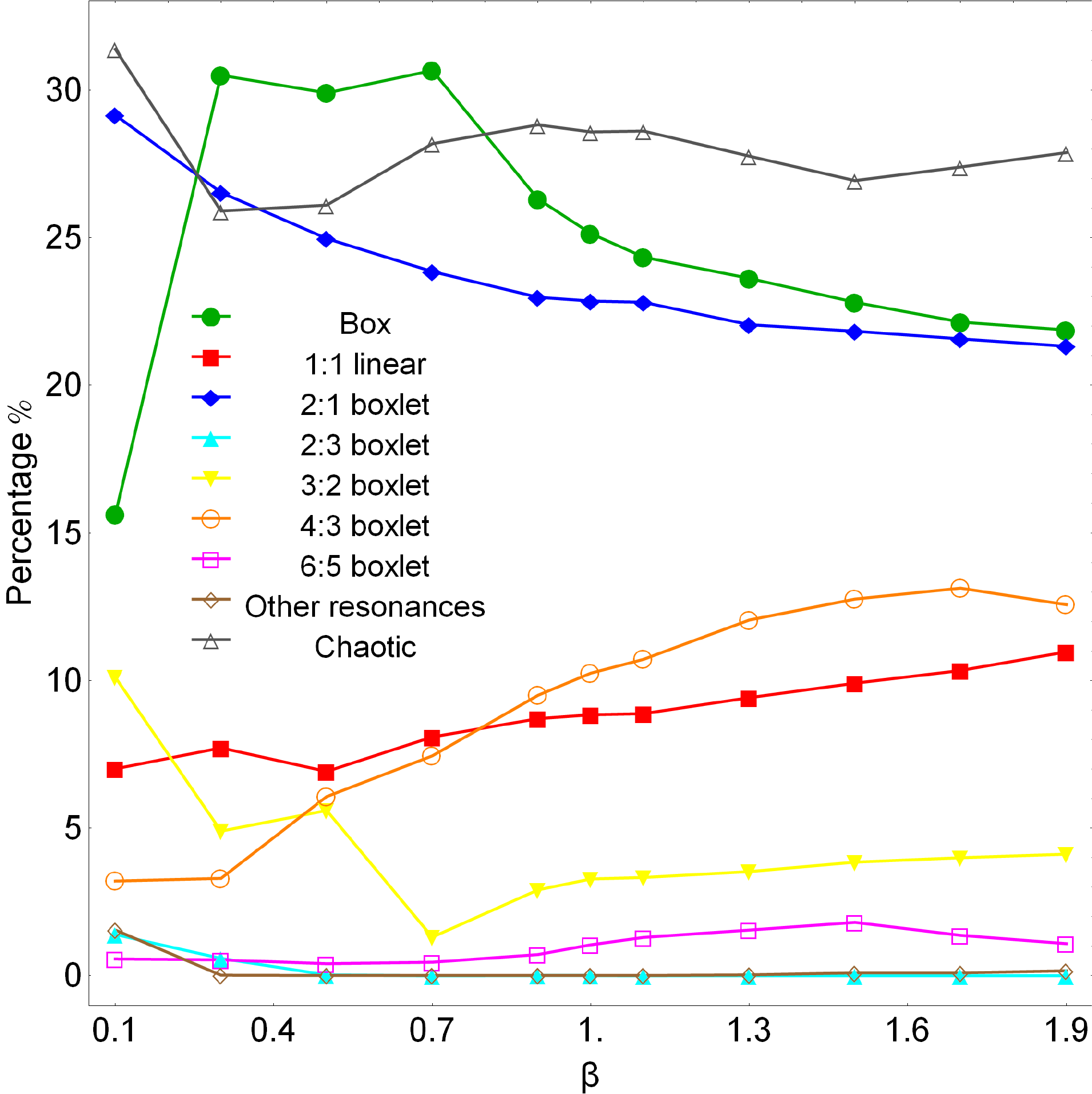}
\caption{Evolution of the percentages of the different kinds of orbits in our galaxy model, when varying the flattening parameter $\beta$ of the central non spherical nucleus.}
\label{percs}
\end{figure}

To study how the flattening parameter $\beta$ of the non spherical central nucleus influences the level of chaos, we let it vary while fixing all the other parameters of our galaxy model. As already said, we fixed the values of all the other parameters and integrate orbits in the meridional plane for the set $\beta = \{0.1,0.3,0.5, ..., 1.9\}$. The exact value of $\beta$ does not influence the total number of initial conditions of orbits per grid. The points in every grid were separated 0.1 units in the $R$ and 0.5 units in the $\dot{R}$ direction, which results to a total of 7173 initial conditions $(R_0,\dot{R_0})$ of orbits under investigation. For the computation of the FLI and SALI of orbits in each grid, we need about 4.5 h of CPU time on a Pentium Dual-Core 2.2GHz PC.

In Figs. \ref{grdprl}(a-f) we present grids of initial conditions $(R_0,\dot{R_0})$ of orbits that we have classified for six values of the flattening parameter $\beta$ when the central nucleus is prolate, that is when $0.1 \leq \beta < 0.9$, or spherical when $\beta = 1$. Here, we can identify all the different regular families by the corresponding sets of islands which are formed in the phase plane. In particular, we see the seven main families already mentioned: (i) 2:1 banana-type orbits surrounding the central periodic point; (ii) box orbits are situated mainly outside of the 2:1 resonant orbits; (iii) 1:1 open linear orbits form the elongated island in the outer parts of the phase plane; (iv) 2:3 resonant orbits corresponding to the set of triple small islands at the outer parts of the grid; (v) 3:2 resonant orbits form the double set of islands above the box orbits; (vi) 4:3 resonant orbits correspond to the inner triple set of islands shown in the phase plane and (vii) 6:5 resonant orbits producing a chain of five tiny islands which are deeply buried in the chaotic sea. Apart from the regions of regular motion, we observe the presence of chaotic areas surrounding all the stability islands. Thus, we may conclude, that when the central nucleus is either prolate or spherical, there is not a unified chaotic domain, at least in this $z = 0$ slice of the phase space. The outermost black thick curve is the ZVC defined by Eq. (\ref{zvc}).

Similar girds of initial conditions $(R_0,\dot{R_0})$ of orbits are given in Figs. \ref{grdobl}(a-f). In this case, we have classified orbits for six values of the flattening parameter $\beta$ when the central nucleus is oblate, that is when $1.1 < \beta \leq 1.9$ (we included again the grid of the spherical model for comparison between prolate and oblate grid models). Apart from the 2:3 resonant family, all the other main families of regular orbits are still the same as in the previous case (prolate nucleus) and form well-defined islands of initial conditions in the grids. We observe, that the structure of the phase plane change very little with increasing $\beta$, unlike the previous case.

Looking carefully at the grids of initial conditions presented in Figs. \ref{grdprl}(a-f) and \ref{grdobl}(a-f) we can draw interesting results regarding the orbital structure of the phase plane which are the following:
\begin{enumerate}
 \item When the central nucleus is prolate with the extreme possible value $(\beta = 0.1)$, we see in Fig. \ref{grdprl}a that apart from the central extended region, the 2:1 resonance occupies also a secondary smaller area at the phase plane.
 \item The 2:3 resonance exits only when $\beta \leq 0.5$. Especially when $\beta = 0.5$ the corresponding islands of the 2:3 resonance are so small, that they appear as lonely points in the grid. Will then see, that for larger values of $\beta$ this resonance becomes highly unstable and therefore disappears from the phase plane.
 \item For $\beta \geq 0.3$ it is evident, that the 3:2 resonance appears itself twice in the grid; mainly above the 2:1 region and inside the box area but also at the outer parts of the box area or even inside the chaotic sea.
 \item Apart from the vast chaotic sea which is present at the outer parts of the phase plane, we can also distinguish a weak chaotic layer inside the box region which embraces the islands of the 3:2 resonance.
 \item When the central nucleus is oblate $(1.1 \leq \beta \leq 1.9)$, the smaller secondary islands of the 3:2 resonance disappear, mainly because the entire area occupied by box orbits which hosts the 3:2 family is being reduced with increasing $\beta$.
\end{enumerate}

Fig. \ref{percs} shows the evolution of the percentages of the chaotic orbits and of the main families of regular orbits as $\beta$ varies. It can be seen, that there is a strong correlation between the percentages of most types of orbits and the value of $\beta$. When the central nucleus is highly prolate $(\beta = 0.1)$ the majority of stars move in chaotic orbits. However, as the nucleus becomes less prolate, the rate of box orbits increases rapidly and box orbits is the most populated family in the interval $0.3 \leq \beta \leq 0.7$. On the other hand, when $\beta > 0.7$ the percentage of chaos increases again and remains the dominant type of motion throughout, fluctuating around 28\%. The percentage of the 2:1 resonant family decreases steadily with increasing $\beta$. On the contrary, both 1:1 and 4:3 resonant families increase gradually their rates mainly at the expense of box orbits and 2:1 banana-type orbits and when $\beta = 1.9$ it seems to tend to a common value around 12\%. Similarly, at the higher value of $\beta$ studied $(\beta = 1.9)$ the percentages of box and 2:1 orbits tend to coincide around 22\%, thus sharing about two fifths of the entire phase plane. The percentage of the 3:2 resonant orbits decreases sharply for small values of $\beta$ $(\beta < 0.7)$, while this tendency is reversed at higher values of the flattening parameter $(\beta \geq 0.7)$. The 2:3 resonant family exists only when $\beta < 0.7$ and holds very small percentages around 1\%. The percentages of the 6:5 and higher resonant orbits are almost unperturbed by the shifting of the value of the flattening parameter $\beta$. Therefore, taking into account all the results obtained from the diagram shown in Fig. \ref{percs}, one may reasonably conclude that the flattening parameter of the central non spherical nucleus affects considerably almost all types of orbits in disk galaxy models.

\begin{figure*}
\centering
\resizebox{0.9\hsize}{!}{\includegraphics{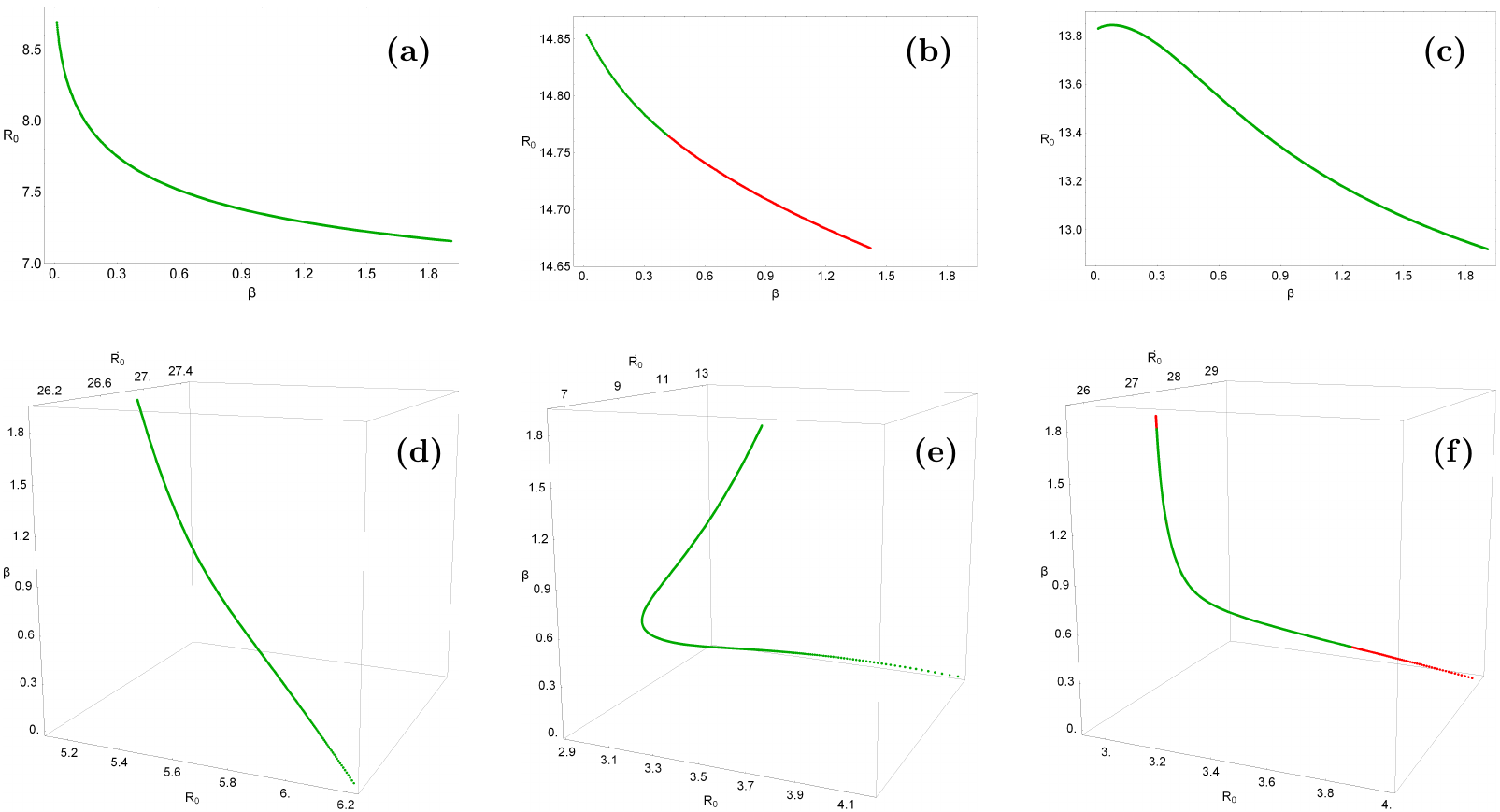}}
\caption{Evolution of the starting position $(R_0,\dot{R_0})$ of the periodic orbits as a function of the flattening parameter $\beta$ of the central non spherical nucleus. (a) 2:1 resonant family; (b) 2:3 resonant family; (c) 4:3 resonant family; (d) 1:1 resonant family; (e) 3:2 resonant family; (f) 6:5 resonant family. Green dots correspond to stable periodic orbits, while unstable periodic orbits are marked with red dots.}
\label{fpo}
\end{figure*}

\begin{figure}
\includegraphics[width=\hsize]{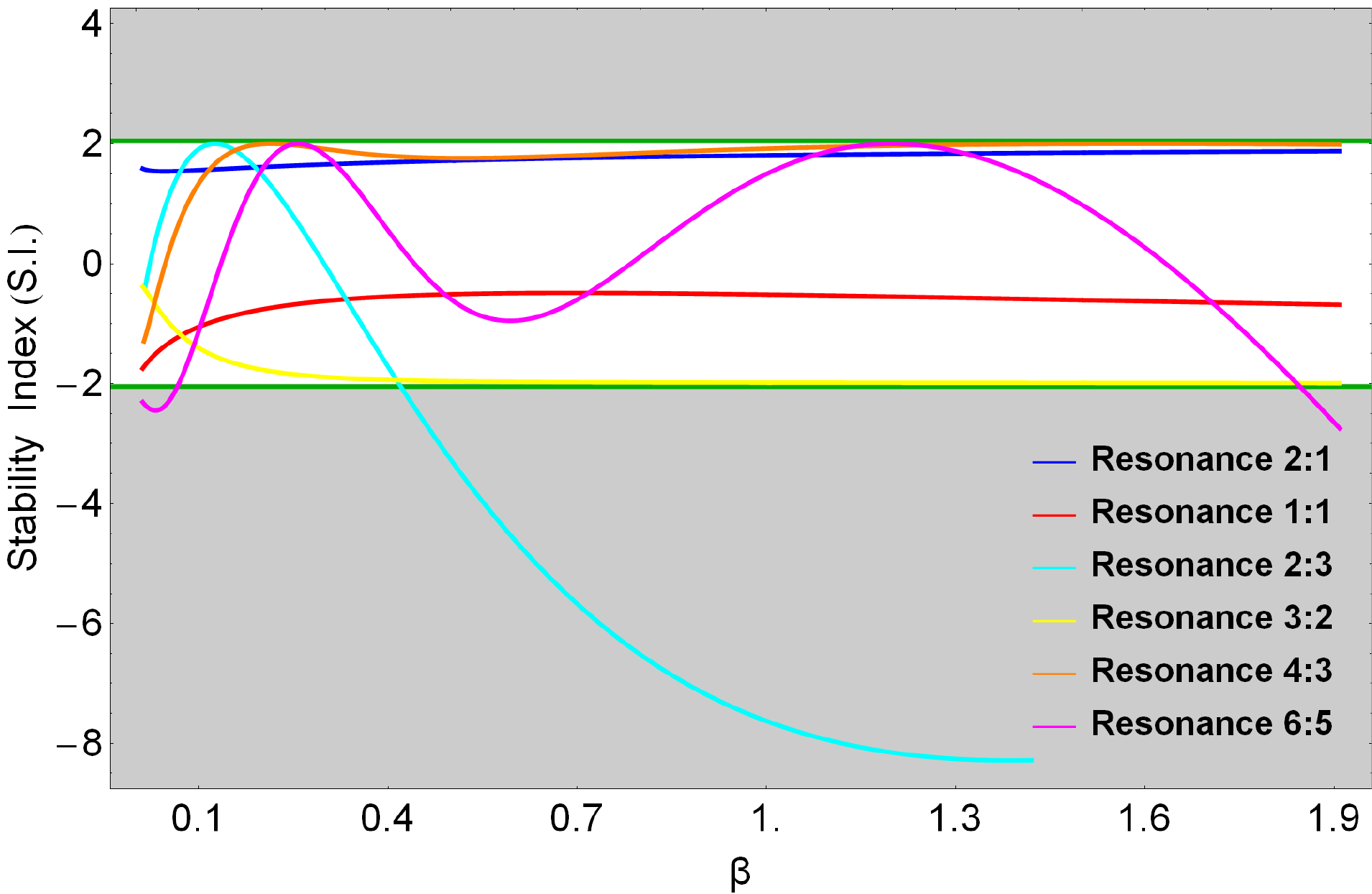}
\caption{Evolution of the stability parameter S.I. of the families of periodic orbits shown in Figs. \ref{fpo}(b-g) as a function of the flattening parameter $\beta$ of the central nucleus. The green horizontal lines at -2 and +2 delimit the range of S.I. for which the periodic orbits are stable. The gray shaded areas, on the other hand, correspond to S.I. values of unstable periodic orbits.}
\label{si}
\end{figure}

Of particular interest, is to investigate how the variation in the flattening parameter of the central non spherical nucleus influences the position of the periodic points of the different families of periodic orbits shown in the grids of Figs. \ref{grdprl} and \ref{grdobl}. For this purpose, we shall use the theory of periodic orbits [\citealp{MH92}] and the algorithm developed and applied in [\citealp{Z13b}]. In Fig. \ref{fpo}(a-f) we present the evolution of the starting position of the parent periodic orbits of the six basic families of resonant orbits. The evolution of the 2:1, 2:3 and 4:3 families shown in Figs. \ref{fpo}(a-c), is two-dimensional since the starting position $(R_0,0)$ of these families lies on the $R$ axis. On the contrary, studying the evolution of the 1:1, 3:2 and 6:5 families of periodic orbits is indeed a real challenge due to the peculiar nature of their starting position $(R_0,\dot{R_0})$. In order to visualize the evolution of these families, we need three-dimensional plots such as those presented in Figs. \ref{fpo}(d-f), taking into account the simultaneous relocation of $R_0$ and $\dot{R_0}$.

The stability of the periodic orbits can be obtained from the elements of the monodromy matrix $X(t)$ as follows:
\begin{equation}
\rm S.I. = {\rm Tr} \left[X(t)\right] - 2,
\end{equation}
where Tr stands for the trace of the matrix, and S.I. is the \emph{stability index}. For each set of values of $\beta$, we first located, by means of an iterative process, the position of the parent periodic orbits. Then, using these initial conditions we integrated the variational equations in order to obtain the matrix $X$, with which we computed the index S.I. Our numerical calculations indicate, that in disk galaxy models with non spherical nuclei, there are stable as well as unstable periodic orbits. In Figs. \ref{fpo}(a-f) green dots correspond to stable periodic orbits, while red dots correspond to unstable ones. We see, that the 2:1, 1:1, 3:2 and 4:3 periodic orbits remain stable throughout the entire range of the values of $\beta$. On the other hand, in the case of the 2:3 resonance there is a limit of stability at $\beta = 0.4151$, while the resonant family ceases to exist when $\beta = 1.4201$. It is worth noticing, that in the grid shown in Fig. \ref{grdprl}c there is no evidence of well-defined islands of initial conditions corresponding to the 2:3 resonance, however, Fig. \ref{fpo}b clearly indicates that the resonance is indeed present, although evidently deeply buried in the chaotic sea due to its unstable nature. Furthermore, one may observe, that the vast majority of the 6:5 periodic orbits are stable, except the regions $0.01 \leq \beta \leq 0.2579$ and $1.8421 \leq \beta \leq 1.9$ in which the periodic orbits become unstable. Thus it becomes clear, that the flattening parameter of the central non spherical nucleus of the disk galaxy plays a fundamental role in the stability of the different regular families, which in turn determines which ones are present in each case.

Another perspective of the stability of the families of periodic orbits is given in Fig. \ref{si}, where we present the evolution of the stability parameter S.I. of the families of periodic orbits shown in Figs. \ref{fpo}(b-g) as a function of the flattening parameter $\beta$ of the central nucleus. The periodic orbits are stable if only the stability parameter (S.I.) is between -2 and +2. We observe, that only the 2:3 and 6:5 families cross the stability boundaries thus entering the gray shaded areas which denote instability. Moreover, we see that the stability index of the 3:2 and 4:3 families evolves almost asymptotically to the stability boundaries nevertheless, it never exceeds them in both cases.

\section{``Fast" indicators of chaos}
\label{fastIndi}

Many methods to determine whether a given orbit is chaotic or regular have been developed in the last decades. The already mentioned mLCN, which is a finite-time estimate of the maximum Lyapunov exponent of an orbit, is generally considered the most robust of all, just because its zero or non-zero value does itself define chaos. However, two common criticisms to this method are the heavy numerical work it involves, and the already mentioned fact that it is not easy to establish a numerical zero to disentangle null mLCNs from non-zero ones, thus sometimes requiring long integration times until an orbit reveals itself as regular or chaotic. This is specially true with sticky orbits, as mentioned before in Section \ref{compmeth}.

Thus, most of the newer methods were developed with a ``fast" performance in mind. Among them, we can mention the Frequency Map Analysis [\citealp{L90}], the stretching numbers, also known as Short Time Lyapunov Characteristic Numbers [\citealp{VC94},\citealp{CV97}], the Fast Lyapunov Indicator (FLI) [\citealp{FGL97}], the Relative Lyapunov Indicator (RLI) [\citealp{SEE00},\citealp{SESF04}], the spectral distance [\citealp{VCE98}], the Mean Exponential Growth factor of Nearby Orbits (MEGNO) [\citealp{CS00},\citealp{MCG11}], the Smaller Alignment Index (SALI) [\citealp{S01},\citealp{SABV04}], the Generalized Alignment Index (GALI) [\citealp{SBA07}], the Fast Norm Vector Indicator (FNVI) [\citealp{Z12d}], the correlation integral [\citealp{CS84},\citealp{C08}], the ``Patterns Method" [\citealp{S06}], the Average Power Exponent (APLE) [\citealp{LVE08}], the Fast Fourier Transform of time intervals between successive points on the PSS [\citealp{KV08}], the dynamical spectra of orbits [\citealp{Z12c}], the method of the low frequency power (LFP) [\citealp{VI92}], the ``0–1" test [\citealp{GM04}], as well as some other more recently introduced techniques [\citealp{H05},\citealp{S05}], among others.

\begin{figure}
\includegraphics[width=\hsize]{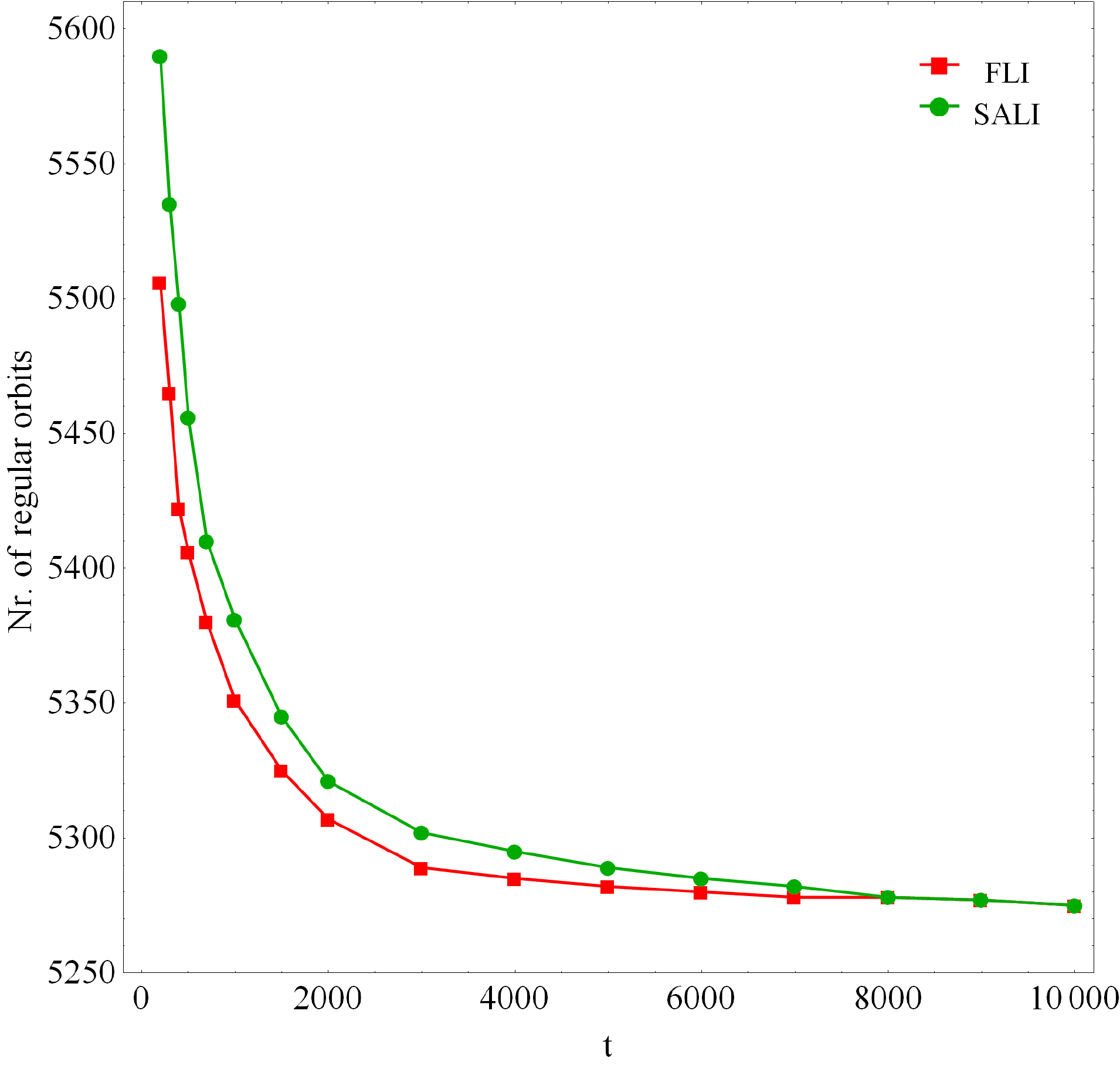}
\caption{Number of orbits classified as regular from a total of 7173 orbits in the $\beta = 1$ model, using the FLI and SALI chaos indicators, as a function of the total time of the numerical integration. For both indicators the threshold values were fixed; 10 for the FLI and $10^{-7}$ for the SALI.}
\label{tfin}
\end{figure}

\begin{figure*}
\centering
\resizebox{\hsize}{!}{\includegraphics{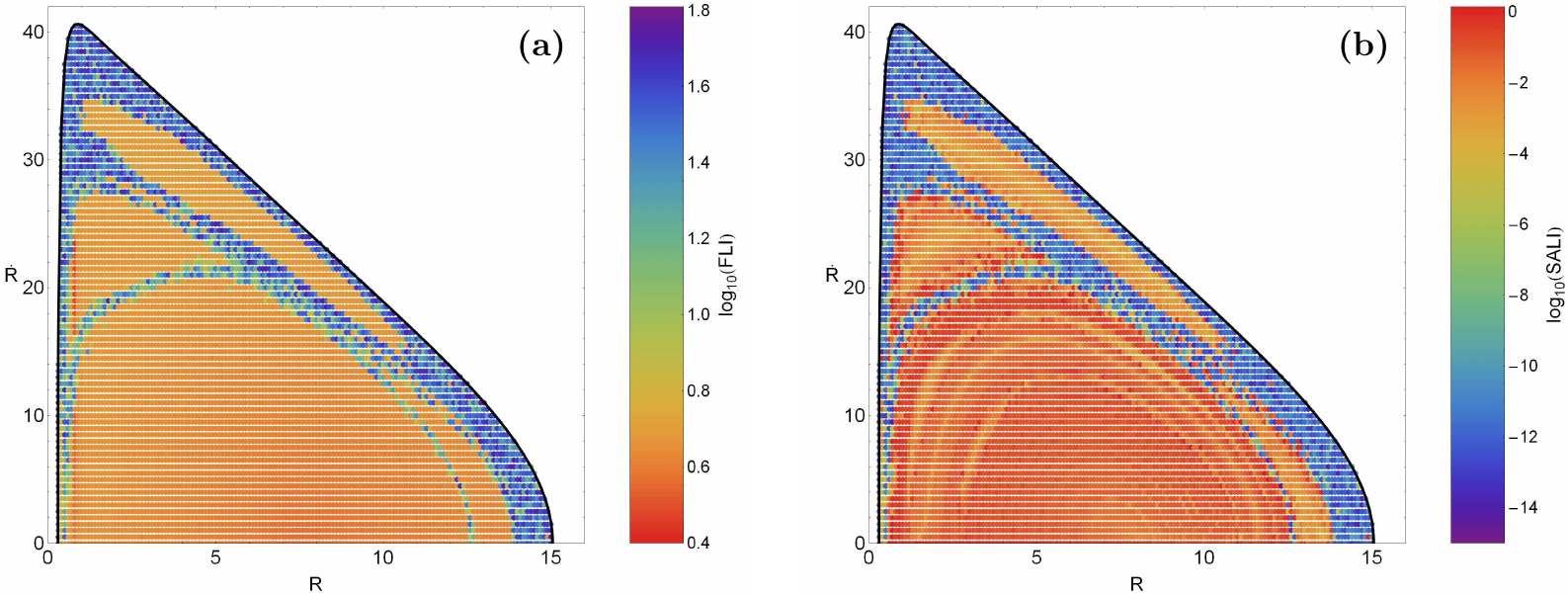}}
\caption{Regions of different values of the (a-left): FLI and (b-right): SALI on the $(R,\dot{R})$ phase plane when the flattening parameter is $b = 1$. Light reddish colors correspond to ordered motion, dark blue/purplr colors denote chaotic motion, while all the intermediate ones suggest initial conditions of sticky orbits, with sticky period more than $10^4$ time units.}
\label{FSgrids}
\end{figure*}

Thus, we had begun our investigation applying both the FLI and SALI indicators using a relatively short integration time of $10^3$ time units, in order to save computation time. But, in an attempt to validate our results, we made a couple of much longer integrations. We were surprised to see, that the classification of a non-negligible number of orbits changed. Fig. \ref{tfin} shows, for the set of 7173 orbits of Fig. \ref{grdprl}f ($\beta = 1$ model), how the number of regular orbits shifted along with the time span of the orbital integration for both chaos indicators. In Fig. \ref{FSgrids} we present two dense grids of initial conditions $(R_0,\dot{R_0})$, where the values of FLI (left) and SALI (right) are plotted using different colors. We clearly observe several regions of regularity indicated by light reddish colors as well as a unified chaotic domain (blue/purple dots), while intermediate colors suggest the location of initial conditions of sticky orbits, with a sticky period more than $10^4$ time units.

\begin{figure*}
\centering
\resizebox{\hsize}{!}{\includegraphics{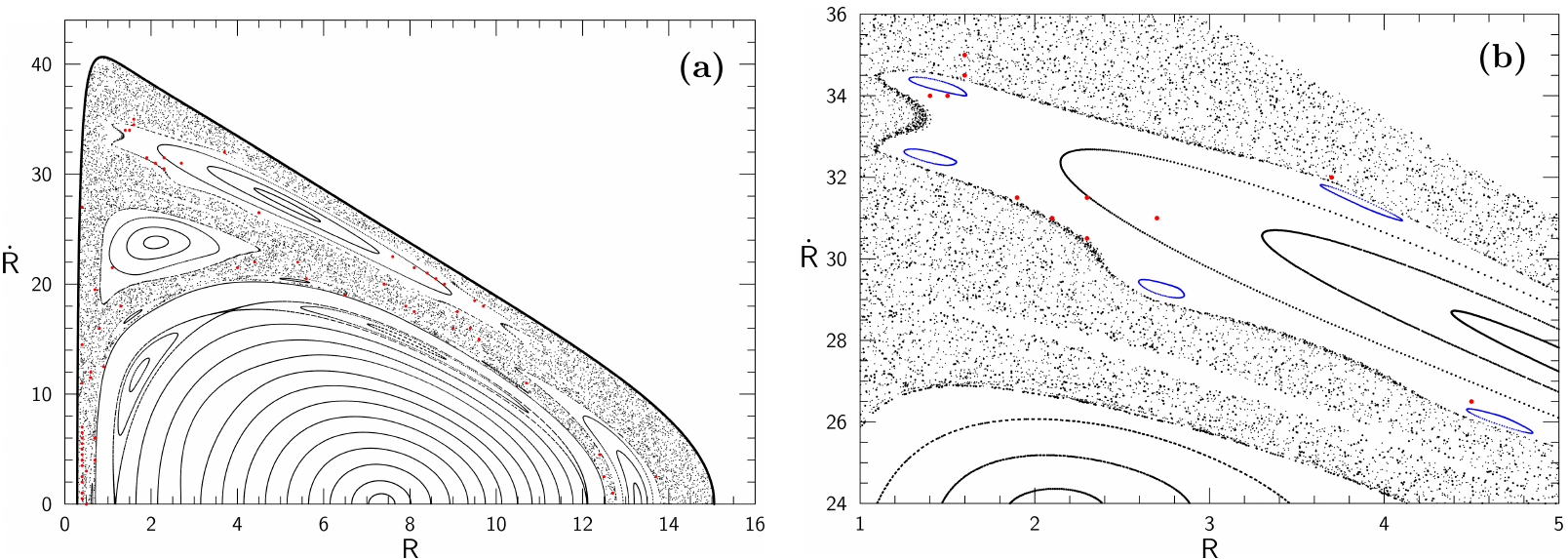}}
\caption{(a-left): The structure of the $(R,\dot{R})$ phase plane when $\dot{R} > 0$ for the $\beta = 1$ model. The red dots correspond to misclassified orbits using together the FLI and SALI with a short integration time of $10^3$ time units. (b-right): Magnification of an area of the PSS shown in Fig. \ref{PSSzm}a which contains tiny hidden stability islands very close to the location of the misclassified orbits.}
\label{PSSzm}
\end{figure*}

In order to see whether the misclassified orbits had some common feature, we plotted in Fig. \ref{PSSzm}a their initial conditions in a $(R, z = 0, \dot{R}, \dot{z} > 0)$ Poincar\'{e} Surface of Section (PSS) (note that it is the classical PSS but with $\dot{R} > 0$). The red dots correspond to orbits that have been misclassified using combined the FLI and the SALI method with an integration of $10^3$ time units. It can be seen, that many of these orbits are located very close to the boundaries of islands of invariant curves, i.e., they are probably sticky orbits. We examined by eye several of these orbits, confirming that they were indeed sticky, behaving as regular ones during a time longer than the shorter integration times.

As per the red dots of Fig. \ref{PSSzm}a which do not fall close to visible islands of invariant curves, we suspected that they are also sticky orbits and therefore, there should be small invisible islands they are stuck to. To prove this, we conducted a more thorough scan of the PSS of Fig. \ref{PSSzm}a. Our investigation was successful, as we may observe in Fig. \ref{PSSzm}b. Thus, most if not all of the misclassified orbits are close to islands of regularity. As a reference, the initial conditions of the orbit producing the chain of the blue islands shown in Fig. \ref{PSSzm}b are: $R_0 = 1.51, z_0 = 0, \dot{R_0} = 34.08$, while the value of $\dot{z_0}$ was obtained from the energy integral (\ref{ham}).

\begin{figure*}
\centering
\resizebox{\hsize}{!}{\includegraphics{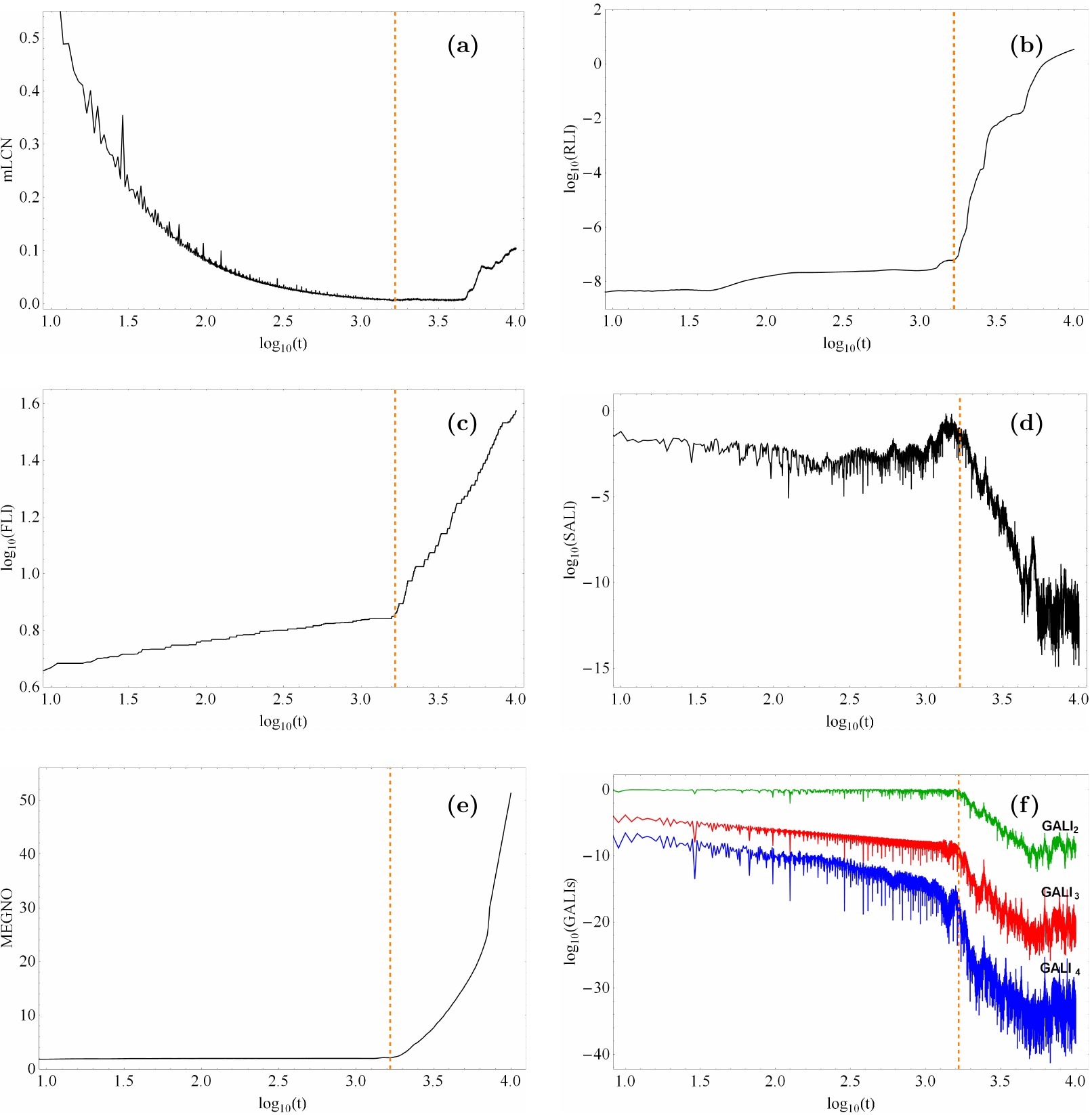}}
\caption{Time evolution of several chaos indicators (a): mLCN, (b): RLI, (c): FLI, (d): SALI, (e): MEGNO and (f): GALI$_k$ for a time interval of numerical integration of $10^4$ time units for a sticky orbit. The vertical orange, dashed line defines the exact time at which the transition from regularity to chaos takes place.}
\label{indices}
\end{figure*}

\begin{figure}
\includegraphics[width=\hsize]{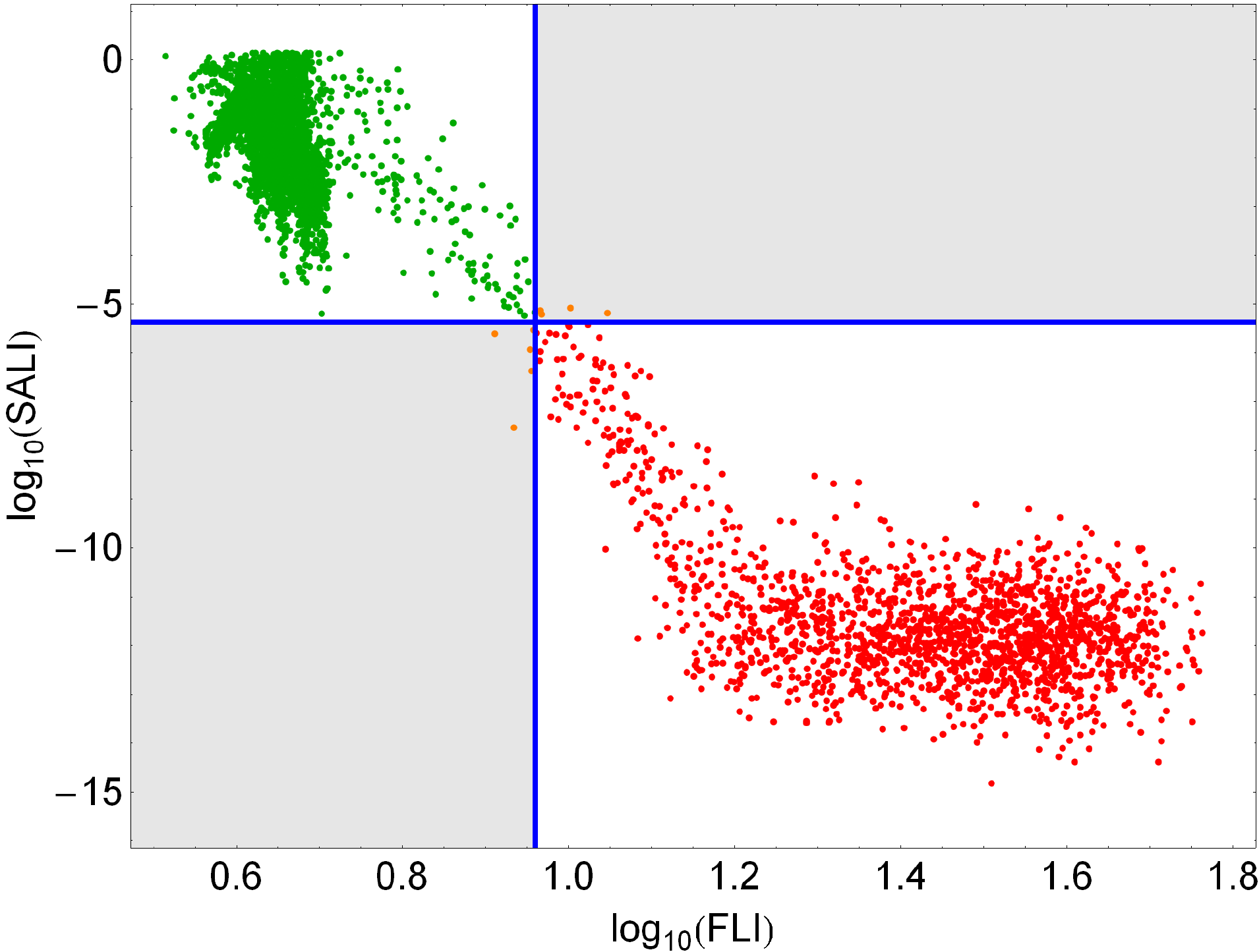}
\caption{FLI and SALI of a set of orbits computed when $\beta = 1$. The blue lines show the resulting thresholds which separate regular (green) from chaotic (red) orbits. The orange dots inside the gray shaded areas, on the other hand, indicate orbits classified differently by both indicators using a time interval of $10^4$ time units of numerical integration.}
\label{trs}
\end{figure}

One may reasonably suppose, that the false classification of some orbits using short integration time may be due to the particular choice of chaos indicators; the FLI and SALI in our case. In order to prove that the false classification is caused by the intrinsic nature of the orbits and therefore is independent of the chaos indicator, we chose a misclassified orbit with initial conditions: $R_0 = 0.4, z_0 = 0, \dot{R_0} = 3.5$ from the $\beta = 1$ model and computed for a time interval of $10^4$ time units not only the FLI and the SALI but also the mLCN, the RLI, the MEGNO and the GALI$_k$. Our results are presented in Figs. \ref{indices}(a-f). We see, that from time-evolution of all chaos indicators we can draw the very same conclusion, that is, for $t \lesssim 1660$ time units this particular orbit seems to be regular, but if we integrate it for a larger time interval ($t > 1660$ time units) the true chaotic nature of the orbit is eventually revealed. Looking carefully at the time-evolution of the mLCN shown in Fig. \ref{indices}a, we observe that even when $t > 1660$ time units the mLCN indicates that the particular orbit is still regular and only after about 3000 time units later (for $t > 4680$) it signifies chaoticity. This result supports our previous statement that sometimes the mLCN requires longer integration times than other chaos indicators until it reveals the true chaotic nature of an orbit.

\begin{table}
   \centering
   \caption{Total number of orbits in each grid that classified differently by both indicators (FLI and SALI) using either short integration time of $10^3$ time units $(N)$ or long integration time of $10^4$ time units $(N')$. Thresholds $T_{\rm F}$ and $T_{\rm S}$ obtained for the FLI and the SALI, respectively from the long integration.}
   \label{table2}
   \begin{tabular}{@{}|l||c|c|c|c|}
      \hline
      $\beta$ & $\log T_{\rm F}$ & $\log T_{\rm S}$ & $N$ & $N'$ \\
      \hline
       0.1 & 0.898 & -5.527 & 29 & 3 \\
       0.3 & 0.871 & -4.892 & 71 & 0 \\
       0.5 & 0.922 & -5.211 & 81 & 7 \\
       0.7 & 0.842 & -5.819 & 37 & 1 \\
       0.9 & 1.011 & -5.418 & 62 & 5 \\
       1.0 & 0.962 & -5.374 & 44 & 8 \\
       1.1 & 0.952 & -5.822 & 35 & 2 \\
       1.3 & 1.019 & -5.013 & 56 & 9 \\
       1.5 & 0.881 & -4.970 & 69 & 0 \\
       1.7 & 0.975 & -5.751 & 41 & 4 \\
       1.9 & 0.981 & -5.406 & 73 & 6 \\
      \hline
   \end{tabular}
\end{table}

We therefore used an integration time of $10^{4}$ time units throughout. We then established new thresholds by taking advantage of our computation of two chaos indicators, as follows. First, the set of orbits of a given grid was integrated, as we already said, for $10^4$ time units (i.e., about $10^{12}$ years, thus avoiding sticky orbits with a stickiness at least of the order of one Hubble time). Then, their FLI and SALI were computed, and we looked for those values of the thresholds that maximised the agreement in the classification of both methods. We found those values which leave less than 1\% of orbits per grid differently classified by both methods. Fig. \ref{trs} shows the resulting thresholds for both indicators when $\beta = 1$. The horizontal and vertical blue lines correspond to the new established threshold values for the FLI and SALI which are $\rm log_{10}(FLI) = 0.962$ and $\rm log_{10}(SALI) = -5.374$. Green dots indicate regular orbits, while chaotic orbits are marked with red dots. However, we can observe eight orange dots, corresponding to orbits for which the classifications of both indicators did not coincide. Similar results apply to all other grid models. As a reference, Table \ref{table2} shows the number of orbits which didn't get the same classification in every model using either $10^3$ or $10^4$ time units of numerical integration. Here we must point out, that all grids contain the same number of initial conditions of orbits which is 7173.

It is evident from Table \ref{table2}, that using larger integration time the number of misclassified orbits in every grid has been considerably reduced. Nevertheless, this don't solve completely the problem; in fact, there will always be (in a non integrable potential) sticky orbits which behave as regular ones during arbitrarily large times, rendering any attempt to develop an algorithm which finds them all in a short time hopeless. Thus, the moral of the story is that the adjective ``fast" which is attached to many chaos detection algorithms should not be interpreted as ``which detects chaos at an earlier dynamical time than the mLCN", as sometimes is used, but as ``computationally faster than the mLCN".

\section{Discussion and Conclusions}
\label{disc}

In the present paper, we used an analytic, axially symmetric galactic gravitational model which embraces the general features of a disk galaxy with an additional central, non spherical and dense nucleus. In order to simplify our study, we chose to work in the meridional plane $(R,z)$, thus reducing three-dimensional motion to two-dimensional. Our aim was to investigate how influential is the flattening parameter $\beta$ which controls the shape of the central nucleus, on the level of chaos and also on the distribution of regular families in our disk galaxy model. Our extensive numerical results suggest, that the level of chaos as well as the different families of regular orbits are indeed very dependent on the shape (prolate, spherical or oblate) of the central galactic nucleus.

A disk galaxy with a non spherical nucleus is undoubtedly a very complex entity and, therefore, we need to assume some necessary simplifications and assumptions in order to be able to study mathematically the orbital behavior of such a complicated stellar system. For this purpose, our model is intentionally simple and contrived, in order to give us the ability to study all the different aspects regarding the kinematics and dynamics of the model. Nevertheless, contrived models can provide an insight into more realistic stellar systems, which unfortunately are very difficult to be studied, if we take into account all the astrophysical aspects. Self-consistent models on the other hand, are mainly used when conducting $N$-body simulations. However, this application is entirely out of the scope of the present research. Once again, we have to point out that the simplicity of our model is necessary; otherwise it would be extremely difficult, or even impossible, to apply the extensive and detailed dynamical study presented in this study. Similar gravitational models with the same limitations and assumptions were used successfully several times in the past in order to investigate the orbital structure in much more complicated galactic systems (e.g., [\citealp{Z12b},\citealp{Z13a}]).

Since a distribution function of the galaxy model was not available so as to use it for extracting the different samples of orbits, we had to follow an alternative path. So, for determining the regular or chaotic nature of motion in our models, we chose, for each set of values of the flattening parameter, a dense grid of initial conditions in the $(R,\dot{R})$ phase plane, regularly distributed in the area allowed by the value of the orbital energy $E$. Each orbit was integrated numerically for a time period of $10^4$ time units (10 billion yr), which corresponds to a time span of the order of hundreds of orbital periods but of the order of one Hubble time. The particular choice of the total integration time was made in order to eliminate sticky orbits (classifying them correctly as chaotic orbits) with a stickiness at least of the order of one Hubble time. Then, we made a step further, in an attempt to distribute all regular orbits into different families. Therefore, once an orbit has been characterized as regular, we then further classified it using a frequency analysis method. This method calculates the Fourier transform of the coordinates and velocities of an orbit, identifies its peaks, extracts the corresponding frequencies and then searches for the fundamental frequencies and their possible resonances.

In this work, we revealed the influence of the flattening parameter $\beta$ of the central non spherical nucleus on the level of chaos and also on the distribution of the regular families among its orbits in disk galaxy models with non spherical nuclei. The main results of our research can be summarized as follows:
\begin{enumerate}
 \item In our galaxy models several types of regular orbits exist, while there is also an extended chaotic domain separating the areas of regularity. In particular, most types of regular orbits, such as the box, 1:1, 2:1, 3:2, 4:3, 6:5, higher resonant and chaotic orbits are always present when $\beta$ varies. The 2:3 resonant family, on the other hand, disappears when the flattening parameter obtains high values $(\beta > 0.5)$. Here we must clarify, that by the term ``higher resonant orbits" we refer to resonant orbits with a rational quotient of frequencies made from integers $> 5$, which of course do not belong to the main families.
 \item There is a strong correlation between the percentages of most types of orbits and the value of the flattening parameter. Almost throughout the values of $\beta$ chaotic motion is the dominant type of motion; only when $0.3 \leq \beta \leq 0.7$ regular motion prevails being the box orbits the most populated family. Generally, the percentages of box and 2:1 resonant orbits decrease, while the rates of the 1:1 and 4:3 resonant orbits exhibit a constant growth with increasing $\beta$. The portion of 3:2 resonant orbits decreases sharply for small values of $\beta$ $(\beta < 0.7)$, while this tendency is reversed at higher values of the flattening parameter $(\beta \geq 0.7)$. The 2:3 resonant family is stable only when $\beta < 0.4$, while for larger values of $\beta$ it becomes unstable and finally ceases to exist. The rates of the 6:5 and higher resonant orbits on the other hand, are almost unperturbed by the shifting of the value of $\beta$. Summarizing, our numerical calculations indicate that the flattening parameter of the central non spherical nucleus affects more or less almost all types of orbits in disk galaxy models.
 \item In disk galaxy models with non spherical nuclei, there are stable as well as unstable periodic orbits. We found, that the 2:1, 1:1, 3:2 and 4:3 periodic orbits remain stable throughout the entire range of the values of $\beta$. On the contrary, the 2:3 and 6:5 families of periodic orbits contain not only stable but also a considerable amount of unstable periodic orbits. Therefore, we may conclude that the flattening parameter $\beta$ affects substantially the stability of the regular families of orbits, hinting at a deep interplay between chaos and proportion of regular families.
\end{enumerate}

Over the last years, several dynamical indicators for distinguishing between ordered and chaotic motion have been developed. The vast majority of these indicators is based on the evolution of sets of deviation vectors in order to characterize an orbit and most of them claim to be equally reliable and fast. Therefore, based on this assumption we applied both the FLI and SALI indicators using a relatively short integration time of only $10^3$ time units, in order to save computation time. However, after performing a few longer integrations using a time interval of $10^4$ time units, we realized that a non-negligible number of orbits was in fact misclassified. What happened was that both chaos indicators failed to identify the true chaotic nature of sticky orbits due to the short available integration time. Thus, we may conclude that the so called ``fast" indicators of chaos cannot reliably identify chaos when the orbits are numerically integrated for short time intervals. They may be computationally faster than the traditional mLCN method, but they cannot overcome the intrinsic difficulty of disentangle sticky orbits form regular ones. As an aside, we found that, by combining two chaos indicators; the FLI and the SALI algorithms in our case, we can define for both methods reliable thresholds, thus separating more accurately chaotic from regular motion. However, the particular thresholds are valid only in our galaxy model. The bottom line of our research is that in every dynamical system one should initially perform thorough statistical tests in order to define the best thresholds of the chaotic indicators and also use sufficient time of numerical integration, thus minimizing as possible the amount of misclassified orbits.

We consider the results of the present research as an initial effort and also a promising step in the task of exploring the orbital structure of disk galaxies with non spherical nuclei. Taking into account that our outcomes are encouraging, it is in our future plans to modify properly our dynamical model in order to expand our investigation into three dimensions, thus unveiling how the flattening parameter influences the nature of the three-dimensional orbits. Also, of particular interest would be to obtain the entire network of periodic orbits, revealing the evolution of the periodic points as well as their stability when varying all the available parameters of our model.

\section*{Acknowledgments}

The authors would like to express their warmest thanks to the two anonymous referees for the careful reading of the manuscript and for all the apt suggestions and comments which allowed us to improve both the quality and the clarity of our paper.

\end{document}